\documentclass[nofootinbib,prx,twocolumn,showpacs,superscriptaddress,notitlepage,superscriptaddress,amsmath]{revtex4-2}

\newcommand\mydots{\hbox to 0.9em{.\hss.\hss.}}
\usepackage{lipsum}  
\usepackage{dsfont}
\usepackage{amsmath}
\usepackage{braket}
\usepackage{amssymb}
\usepackage{amsthm}
\usepackage{algpseudocode}
\usepackage{algorithm}
\usepackage{amsfonts}
\usepackage{comment}
\usepackage[normalem]{ulem}
\usepackage{graphicx}
\usepackage{color,framed}
\usepackage[colorlinks=true,citecolor=blue,linkcolor=blue,urlcolor=blue]{hyperref}
\usepackage{enumerate}
\usepackage{lipsum}
\usepackage{slashed}
\usepackage{url}
\usepackage{bbm}
\usepackage{chngcntr}
\counterwithout{equation}{section}
\usepackage{tikz,pgfplots}
\usepackage{pgfplotstable}
\usepackage{siunitx}
\usepackage{comment}
\usepackage{graphicx}
\usepackage{multirow}

\makeatletter
\newcommand{\setlabel}[1]{\edef\@currentlabel{#1}\label}
\makeatother

\usepgfplotslibrary{fillbetween}

\hypersetup{
    colorlinks=true, 
    linktoc=all,     
    linkcolor=blue,  
}

\begin{document}

\title{Experimental Demonstration of Break-Even for the Compact Fermionic Encoding}
\author{Ramil Nigmatullin} \author{Kevin Hemery} \author {Khaldoon Ghanem}
\affiliation{Quantinuum, Leopoldstrasse 180, 80804 Munich, Germany}

\author{Steven Moses}
\author{Dan Gresh}
\author{Peter Siegfried}
\author{Michael Mills}
\author{Thomas Gatterman}
\author{Nathan Hewitt}
\affiliation{Quantinuum, 303 S Technology Ct, Broomfield, CO 80021, USA}

\author{Etienne Granet}
\author{Henrik Dreyer}
\thanks{henrik.dreyer@quantinuum.com}
\affiliation{Quantinuum, Leopoldstrasse 180, 80804 Munich, Germany}
\date{\today}

\begin{abstract}
The utility of solving the Fermi-Hubbard model has been estimated in the billions of dollars \cite{agrawal_quantifying_2024}. Digital quantum computers can in principle address this task, but have so far been limited to quasi one-dimensional models \cite{stanisic_observing_2022, hemery_measuring_2024, arute_observation_2020}. This is because of exponential overheads caused by the interplay of noise and the non-locality of the mapping between fermions and qubits. Here, we show experimentally that a recently developed local encoding \cite{derby_compact_2021} can overcome this problem.
We develop a new compilation scheme, called ``corner hopping",  that reduces the cost of simulating fermionic hopping by 42\% which allows us to conduct the largest digital quantum simulations of a fermionic model to date, using a trapped ion quantum computer to prepare adiabatically the ground state of a $6\times6$ spinless Fermi-Hubbard model encoded in 48 physical qubits. We also develop two new error mitigation schemes for systems with conserved quantities, one based on local postselection and one on extrapolation of local observables.
Our results suggest that Fermi-Hubbard models beyond classical simulability can be addressed by digital quantum computers without large increases in gate fidelity.
\end{abstract}

\maketitle

\section{Introduction}

The Hubbard model plays a central role in interacting quantum materials. On the one hand, more than sixty years of progress have revealed evidence of the rich physics of the model, ranging from antiferromagnetism and charge density waves to superconductivity and quantum spin liquids~\cite{arovas_hubbard_2022,qin_hubbard_2022}, while, in parallel, more and more materials have been discovered in which many of the same phases can be observed experimentally~\cite{wilson_charge-density_1974,bednorz_possible_1986,takada_superconductivity_2003,catalano_rare-earth_2018, wu_hubbard_2018}. On the other hand, even though numerical methods are starting to become reliable for the ground state physics~\cite{ponsioen_period_2019, xu_coexistence_2024}, many questions remain unsolved, especially about the dynamics of the model~\cite{ray_photo-induced_2024, zhang_photoinduced_2022, kaneko_photoinduced_2019, white_correlations_2019, mehio_hubbard_2023, fava_magnetic_2024}. This is one of the major reason for the tremendous effort that has been devoted to the construction of analog quantum simulators~\cite{mitra_quantum_2018,bakr_quantum_2009, greif_short-range_2013, hilker_revealing_2017, mazurenko_cold-atom_2017}. Digital quantum computers are well-suited to complement analog devices for this task due to their ability to simulate arbitrary Hamiltonians on demand and to allow for quantum error correction which may eventually allow them to reach lower temperatures. Still, the capabilities of digital quantum computers for fermionic simulations lag far behind that of analog simulators, with experiments restricted to quasi one-dimensional geometries \cite{stanisic_observing_2022, hemery_measuring_2024, arute_observation_2020}.

\begin{figure}[ht!]
\begin{center}
\includegraphics[width=0.45\textwidth]{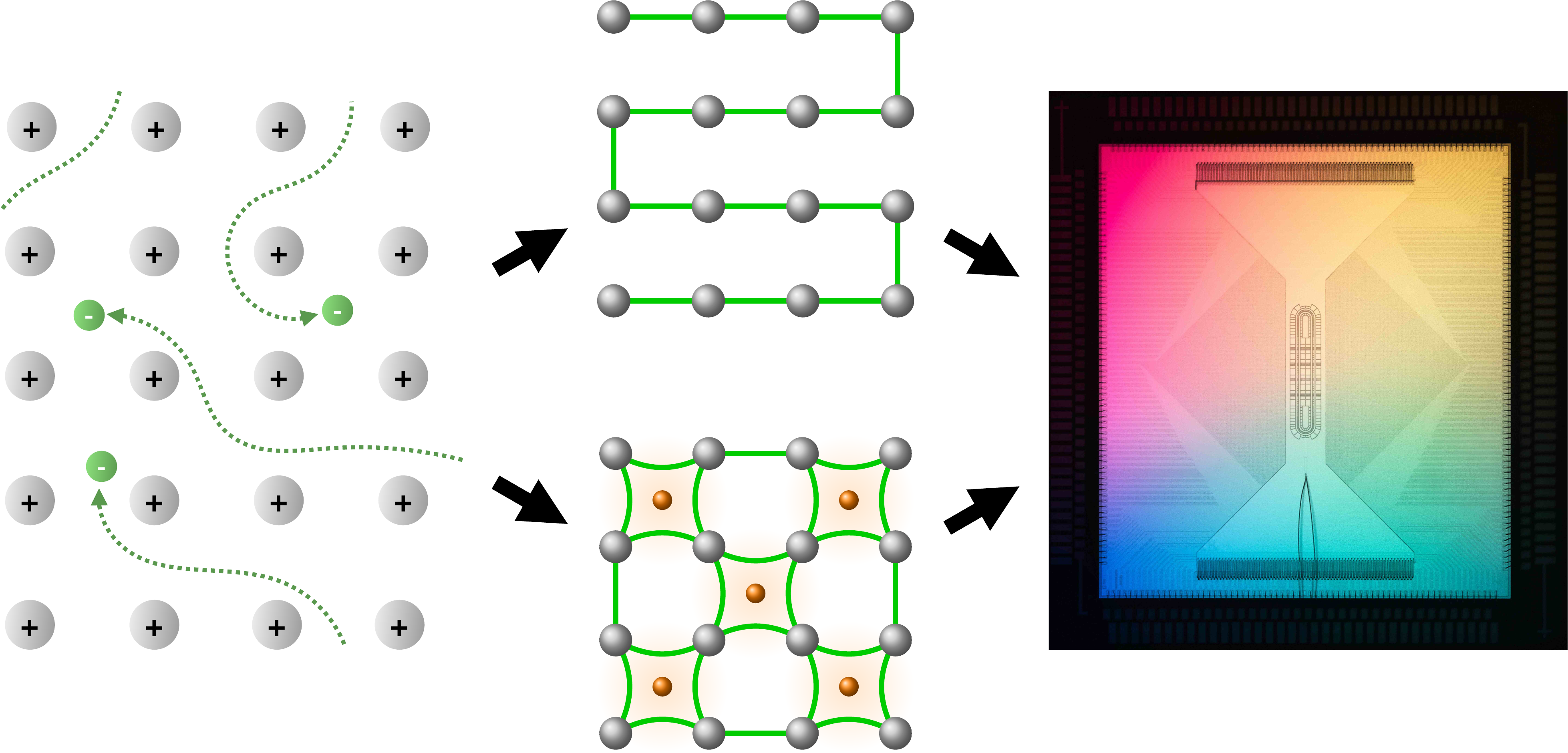}
\caption{\textbf{Conceptual Figure.} Fermions, like electrons in a solid, can be simulated by qubits in one of two ways: one can either unravel the material into a one-dimensional chain and simulate a non-local model, or embed the material into a larger, topologically ordered system that stores the fermionic exchange statistics in a local way, at the cost of introducing ancillae. Here, we provide experimental data from a trapped ion quantum computer that the latter is competitive with the former and has a scaling advantage which may be exponential.
\label{fig_conceptual}}
\end{center}
\end{figure}

\begin{figure*}[!ht]
	\centering
\includegraphics[width=1.0\textwidth]{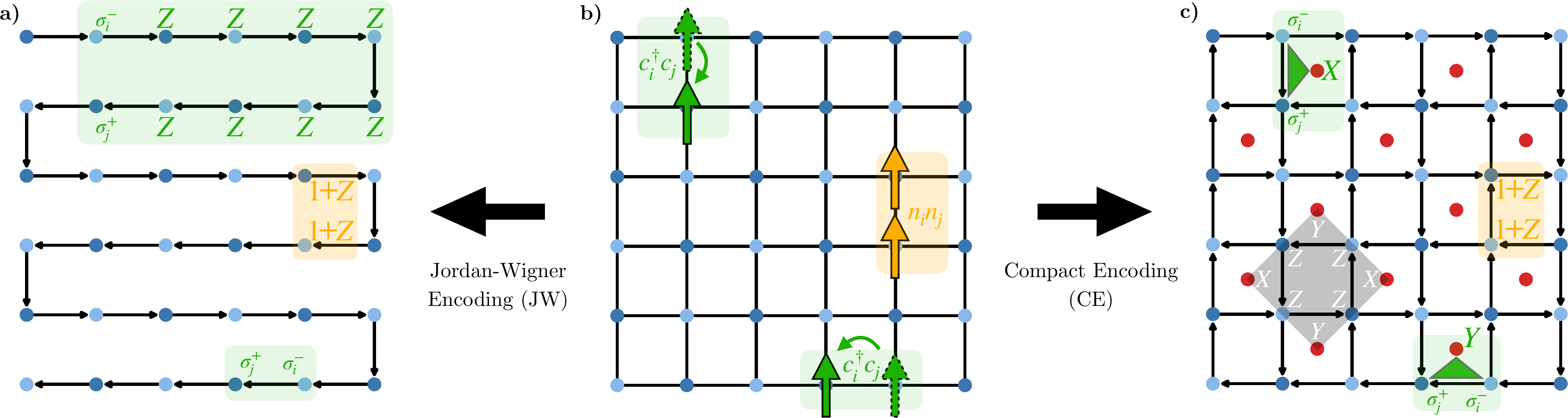}
\caption{\textbf{The Fermi-Hubbard model and the two fermion-to-qubit mappings implemented in this work.} (b) The spinless Fermi-Hubbard model we consider has two competing terms: a nearest-neighbour hopping term (green) that prefers fermions to delocalise across the system, and an interaction term (yellow) which repels particles residing on neighbouring lattice sites. We initialise the state with fermions on the dark sites of the checkerboard pattern and attempt to adiabatically lower the energy of this state. a) The Jordan-Wigner Encoding unravels the system into a one-dimensional chain. Hopping terms are implemented with qubit operators whose weight grows with their distance on this chain. c) The Compact Encoding stores the fermionic exchange statistics in topologically ordered ancillae (red), such that all qubit hopping operators have weight 3. The stabilisers (shaded gray) are set to +1 by initialising a toric code state and are conserved throughout the evolution. \label{fig1}}
\end{figure*}

Recent theoretical work sheds light on the extreme difficulty of simulating proper two-dimensional fermionic systems with a qubit-based quantum computer: Generically, computing local expectation values $\braket{M}$ on a noisy quantum computer incurs a signal-to-noise ratio (SNR) scaling as
    \begin{align}
        \mathrm{SNR} := \frac{\braket{M}_\mathrm{noisy}}{\braket{M}_\mathrm{noiseless}} = \mathcal{O}\left( e^{-\varepsilon \chi T} \right)
        \label{eq_snr}
    \end{align}
where $\varepsilon$ is the error per qubit per time step, $\chi$ is a sensitivity parameter (which may depend on system size) and $T$ is the number of time steps. To encode the electronic into qubit degrees of freedom, previous experiments have focused on the Jordan-Wigner (JW) encoding \cite{jordan_uber_1928}. Taking as an example a two-dimensional Hubbard model on a square lattice with $L \times L$ sites, the best known compilation schemes for this encoding require $\mathcal{O}(L)$ gates per qubit per Trotter step \cite{kivlichan_quantum_2018}, leading to $\varepsilon \propto L$, even if gate error is held constant with increasing qubit number. More subtly, it was shown that, for local Hamiltonian simulation, $\chi \propto \mathrm{weight}(M)$, the average number of non-identity components in the Pauli-string expansion of the (traceless) observable $M$~\cite{granet_dilution_2024} (and this insensitivity of local observables to errors has been found to appear in both energy-conserving~\cite{chertkov_exploring_nodate} and adiabatic dynamics~\cite{schiffer_quantum_2024}). The average weight of a nearest-neighbour hopping operator in the JW mapping becomes $\mathcal{O}(L)$, implying $\chi \propto L$ as well. The JW encoding thus suffers from both a problem of susceptibility (extensive number of gates per Trotter step) as well as a problem of sensitivity (if an error occurs it can affect extensively many observables) to noise. These two effects conspire to attenuate the signal as $\mathrm{SNR} = \mathcal{O}(\exp(-L^2T))$. Since $\mathrm{SNR}^{-2}$ is linearly related to the minimum number of shots required to mitigate errors up to a desired accuracy, even the fastest quantum computer will eventually succumb to this exponential roadblock. 

In this work, we show experimental evidence that a recently proposed alternative to the JW encoding~\cite{derby_compact_2021, jafarizadeh_recipe_2024} can overcome this roadblock. This so-called ``Compact Encoding" (CE) has previously fallen short of JW-based implementations due to constant overheads associated with simulating the hopping of electrons through the lattice. We introduce a new CE compilation scheme that reduces the gate cost of simulating hopping on the square lattice by $42\%$.  This allows us to implement both encodings on Quantinuum's H2 trapped-ion quantum computer and show that the CE significantly  outperforms JW when adiabatically preparing the ground state of a spinless Fermi-Hubbard model on $6 \times 6$ sites. Furthermore, in the CE, the raw data automatically comes with information about a subset of conserved quantities (stabilisers) which we use to develop two local error mitigation schemes for the CE that allows one to keep a much larger fraction of bits compared to global schemes like particle number filtering.

\section{Model, Algorithm \& Encodings}
\label{sec_model}
\begin{figure*}[!ht]
	\centering
\includegraphics[width=1.0\textwidth]{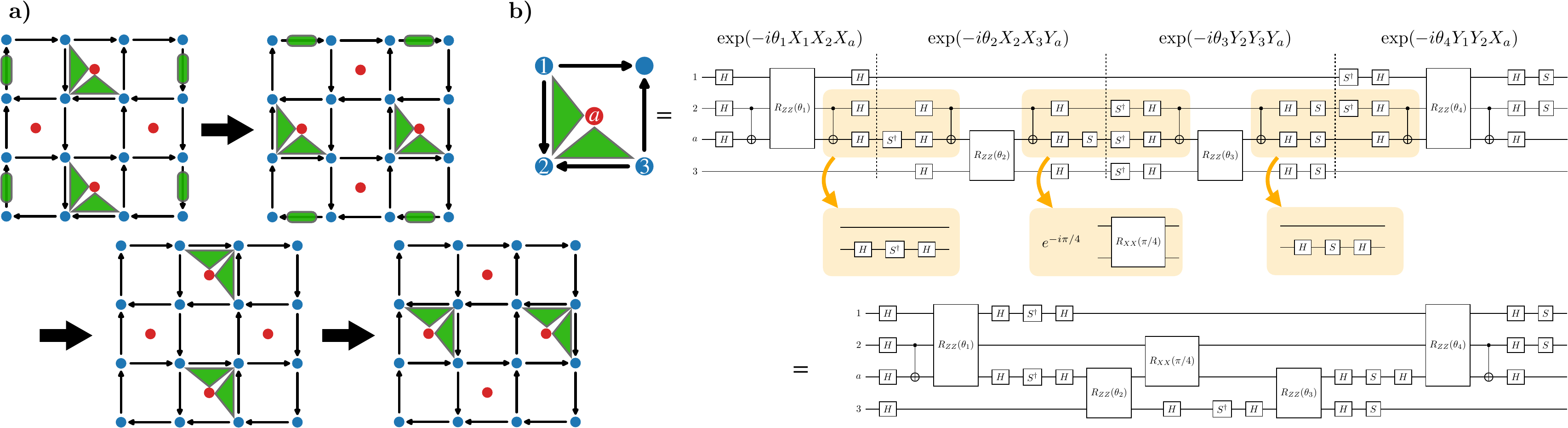}
\caption{\textbf{``Corner Hopping" compilation scheme for simulating the hopping operator~\eqref{eq_unitaries} in the compact encoding.} (a) We group all three-body hopping terms into four sets of corners, grouped by sublattice and by whether the corner is converging (arrows pointing inwards) or diverging (arrows pointing outwards), here shown on a $4 \times 4$ lattice. Hopping at the edges can be implemented in parallel. (b) Each hopping around a corner contains four three-body-terms, which would be compiled via Pauli gadgets into 12 two-body gates (4 partial and 8 maximal entanglers). The circuit identities in the orange boxes reduce this count to 7 (4 partial and 3 maximal entanglers). The decomposition holds for any value of $\theta_1, \dots \theta_4$. In practice, $\theta_j = \pm \tau t/2$, depending on whether the corner is converging or diverging. \label{fig2}} 
\end{figure*}

\begin{figure*}[!t]
	\centering
\includegraphics[width=1.0\textwidth]{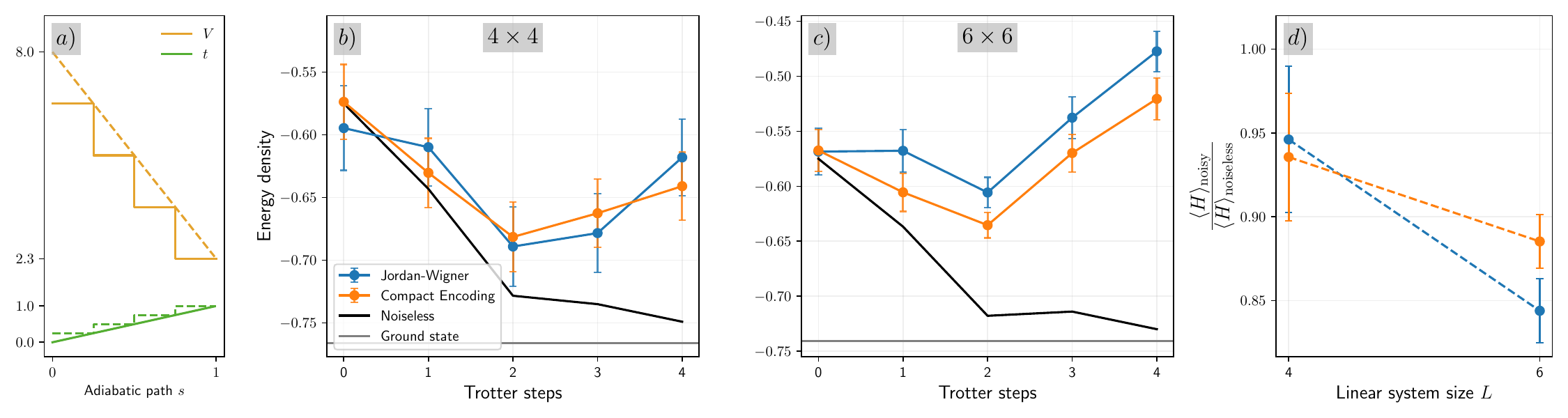}
\caption{\textbf{Raw Data for the Adiabatic Evolution~\eqref{eq_adiabatic_evolution}.} a) Parameter setting for the adiabatic evolution here for $T=4$ Trotter steps. b,c) Experimental results on the energy density with respect to~\eqref{eq_hamiltonian} on the $4 \times 4$ and $6 \times 6$ lattice. d) Signal-to-noise ratio at 2 Trotter steps as a function of system size. All points are raw data to which no error mitigation has been applied. Ground state refers to the ground state energy in the half-filled particle number sector. All adiabatic evolutions are complete. As shown in section~\ref{sec_trotter_error_encodings}, the difference between JW and CE for the noiseless case is negligible in comparison with the error bars.
\label{fig_main_result}}
\end{figure*}

To compare the performance of different fermionic encodings, we focus on the following representative task: The goal is to prepare an approximate ground state in the half-filled sector of a single-species Hubbard model (also known as the $t-V$-model) on $N=L \times L$  square lattices ($L$ even) with open boundary conditions:
    \begin{align}
        H = -t \sum_{\braket{ij}} \left( c_i^\dagger c_j + c_j^\dagger c_i\right) + V \sum_{\braket{ij}} \left(n_i n_j-\frac{1}{4}\right) \label{eq_hamiltonian}
    \end{align}
Here, $t=1$ describes nearest-neighbour hopping and $V=2.3$ is a strong, repulsive Coulomb interaction between nearest-neighbours. To this end, we initialise the system in a state $\ket{\psi}$ that contains fermions in a checkerboard pattern, as shown in Fig.~\ref{fig1}b, and employ an adiabatic algorithm to effectively cool the system. That is, we implement a time evolution under Hamiltonian~\eqref{eq_hamiltonian}, but with a linear ramp in the parameters $V(s) = V_i - s (V_i-V)$ with $V_i = 8.0$, and $t(s)=s$, where $s \in [0,1]$, cf. Fig~\ref{fig_main_result}a. The adiabatic evolution is split into $T$ first order Trotter steps, i.e. 
\begin{equation}
\label{eq_adiabatic_evolution}
    U = U(1) \, U\left(\frac{T-1}{T}\right)\dots U\left(\frac{1}{T}\right),
\end{equation}
where $U(s) = U_\mathrm{int}(s) U_\mathrm{hop}(s)$ and
\begin{equation}
\label{eq_unitaries}
\begin{aligned}
    U_\mathrm{hop}(s) &= e^{i \tau t(s) \sum_{\braket{ij}} \left( c_i^\dagger c_j + c_j^\dagger c_i\right)} \\
    U_\mathrm{int}(s) &=e^{-i \tau V(s) \sum_{\braket{ij}} n_i n_j }
\end{aligned}
\end{equation}
where we are using $\tau = 0.2$. The quality of the state is assessed by computing the energy density per bond, $e = \braket{\psi|U^\dagger H U|\psi}/2L(L-1)$.

This is the testbed in which now we want to assess the scaling performance of two fermionic encodings, when executed on a quantum computer. To specify such an encoding, one must define qubit operators which preserve the commutation relations of the operators appearing in in~\eqref{eq_hamiltonian} and~\eqref{eq_unitaries} as well as a qubit state corresponding to the initial state $\ket{\psi}$. The JW encoding is shown in Fig.~\ref{fig1}a. A one-dimensional ordering of the fermionic modes is chosen which allows the mapping
\begin{equation}
\begin{aligned}
    c_i^\dagger c_j &\overset{\mathrm{JW}}{\longrightarrow} \sigma_i^+ Z_{i+1} \dots Z_{j-1} \sigma^-_j \\
    n_i n_j &\overset{\mathrm{JW}}{\longrightarrow} \frac{1-Z_i}{2} \frac{1-Z_j}{2},
\end{aligned}
\end{equation}
where $2\sigma^{\pm} = X \pm iY$, to be well-defined. The average weight of the Pauli strings corresponding to interaction, horizontal, and vertical hopping terms is $2$, $2$, and $L+1$, respectively. We implement time evolution under the interaction terms~\eqref{eq_unitaries} using a depth-4 circuit of $\exp(-i \tau V Z_i Z_j)$-gates. The hopping terms do not mutually commute and must be further Trotterised. To this end, we use the fermionic swap network shown in~\cite{cade_strategies_2020}, see~\ref{sec_fSWAP_networks} for details. To measure the hopping energy at the end of the circuit, we partition the bonds into four commuting sets and diagonalise the hopping operator $(XX + YY)/2$~\cite{cade_strategies_2020}. The interaction energy is obtained by collapsing the state in the $Z$-basis. In the quantum computer that we are using, SWAP gates can be implemented by physically moving the ions around and thus the total number of two-qubit $\exp(-i\theta/2 Z_i Z_j)$-gates for one Trotter layer for $L$ even is $L^3+3 L^2 - 4L$ for $U_\mathrm{hop}$ and $2L(L-1)$ for $U_\mathrm{int}$ (cf. section \ref{sec_results}).

The operating principle of the compact encoding is rather different (Fig.~\ref{fig1}c). The introduction of ancillae (red) allow for the storage of the fermionic parity information in a local way:
\begin{equation}
\begin{aligned}
    c_i^\dagger c_j &\overset{\mathrm{CE}}{\longrightarrow} \sigma_i^+ P_a \sigma^-_j \\
    n_i n_j &\overset{\mathrm{CE}}{\longrightarrow} \frac{1-Z_i}{2} \frac{1-Z_j}{2},
\end{aligned} \label{eq:CE_mapped}
\end{equation}
where $a$ is the ancilla adjacent to bond $\braket{ij}$, $P$ is the Pauli-$Y$ matrix for horizontal bonds and $+X$ ($-X$) for down-pointing (up-pointing) vertical bonds. $P_a$ is the identity at the edges of the open system, where no ancilla is nearby. All operators are now local with the average weight of the Pauli strings corresponding to interaction and hopping terms being $2$ and $3-1/(L-1)$, respectively. While $U_\mathrm{int}$ is implemented in exactly the same way as in the JW encoding, simulating $U_\mathrm{hop}$ is more subtle. This is because three-body operators like $\exp(i \tau XX P_a)$ are not natively available on the quantum computer we are targeting, and hence must be decomposed. The standard construction based on Pauli gadgets compiles one such term into 3 two-qubit gates on an architecture where partial entanglers are available (4 if not)~\cite{nielsen_quantum_2010}. The total two-qubit gate count is then $12 L^2 - 20L$. However, as shown in Fig.~\ref{fig2}, by grouping hopping terms in a suitable way, we can reduce the cost per two bonds across a corner from 12 to 7 two-qubit gates, reducing the overall two-qubit count for $U_\mathrm{hop}$ to $7L^2-10L$. This new compilation scheme moves the cross-over in terms of two-qubit gate cost from $L \approx 10$ to $L \approx 4$, i.e., into the regime that is experimentally accessible on the device that we are using (cf. Fig.~\ref{fig_gatecounts}). The energy is measured by collapsing the qubits locally in the bases shown in section~\ref{sec_state_prep_and_measurement}.

\section{Experimental Results}
\label{sec_results}
To study the scaling of the effect of noise on both JW and CE, we execute the circuits described in the previous section, corresponding to adiabatic evolution for $T \in \{1,2,3,4\}$ Trotter steps on lattices of size $4\times 4$ and $6\times 6$, on Quantinuum's H2 quantum computer. Quantum information in this device is stored in hyperfine levels of 56 ${}^{171}\mathrm{Yb}^+$ ions. The dominant source of error are two-qubit gate errors $\varepsilon(\theta) \approx (2.9 \, \theta/\pi + 0.46)\times10^{-3}$ for the native entangling gate $\exp(-\theta/2 \, Z_i Z_j)$. Smaller but non-negligible is the impact of so-called memory error, which can be ascribed to coherent dephasing $\exp(-i \phi Z)$ of the qubit states while idling. Other sources of error, like those coming from single-qubit gates are negligible in comparison. Crucially, since gate operations are carried out in spatially separated gate zones, the probability of one gate operation adversely affecting extensively many qubits is negligibly small. Therefore, the only errors that depend on system size are memory errors, since they scale with the total shot time which is proportional to not only circuit depth but also system size. Detailed characterisations of the machine have been carried out~\cite{moses_race-track_2023, decross_computational_2024}.

Our main results are shown in Fig.~\ref{fig_main_result}. For both settings, we see a non-monotonous behavior of the energy with time. While the experimental data initially closely follows the adiabatic cooling of the noiseless algorithm, there is an inflection point at which errors start heating the state. The non-trivial minimum, which is specific to the noise in the quantum computer, occurs after two Trotter steps in all of our experiments. For clarity, we reiterate that the adiabatic evolution is complete for each $T$, e.g. $T=2$ does not mean a $T=4$ protocol that is prematurely aborted.

On the $4 \times 4$-lattice, both JW and CE follow a similar trajectory and our data does not allow us to make statements about their relative performance at this system size. A clear trend develops, however, on the larger $6 \times 6$ system: Here, the probability that the Compact Encoding outperforms the Jordan-Wigner encoding on at least two of the four non-trivial data points is greater than $99\%$. The likelihood that CE produces the overall lowest-energy state at two Trotter steps is greater than $88\%$. We can put the sensitivity to noise into a scaling perspective. Focusing on the lowest energy  circuit at two Trotter steps, the SNR shown in Fig~\ref{fig_main_result}d, shows a much more pronounced dependence on system size for JW over CE.

\begin{figure*}[!ht]
	\centering
\includegraphics[width=1.0\textwidth]{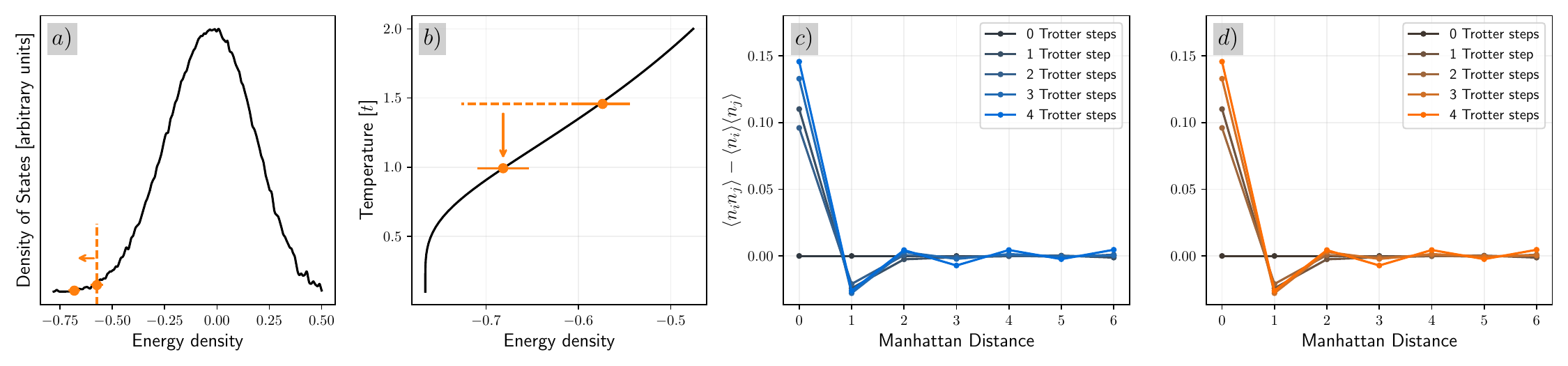}
\caption{\textbf{Physical properties of the prepared state.} a) The spectrum of~\eqref{eq_hamiltonian} on a $4 \times 4$ lattice in the half-filled sector. The quantum computer is initialised in the mean-field ground state in that sector and evolved adiabatically to a lower energy, here showing the results of the Compact Encoding. A Gaussian broadening with width $\sigma = 0.005$ has been applied to the classically obtained spectrum. b) Temperature of the canonical ensemble in the half-filled sector as a function of energy in units of the hopping parameter $t$.  c) Charge density correlations generated by the Jordan-Wigner encoding on the $6 \times 6$ lattice. Despite the net energy increasing after two Trotter steps, staggered charge density correlations are more pronounced after longer evolutions.  d) Same quantity as c), but for the Compact Encoding. All points are raw data to which no error mitigation has been applied. 
Average (max) standard error on the mean for distance 0 is 0.0057 (0.0083) for Jordan-Wigner and 0.0062 (0.0087) for Compact Encoding and 0.0026 (0.0046) for Jordan-Wigner and 0.0027 (0.0048) for the Compact Encoding for distance $>0$. \label{fig_physics}}
\end{figure*}

\begin{figure*}[!ht]
	\centering
\includegraphics[width=1.0\textwidth]{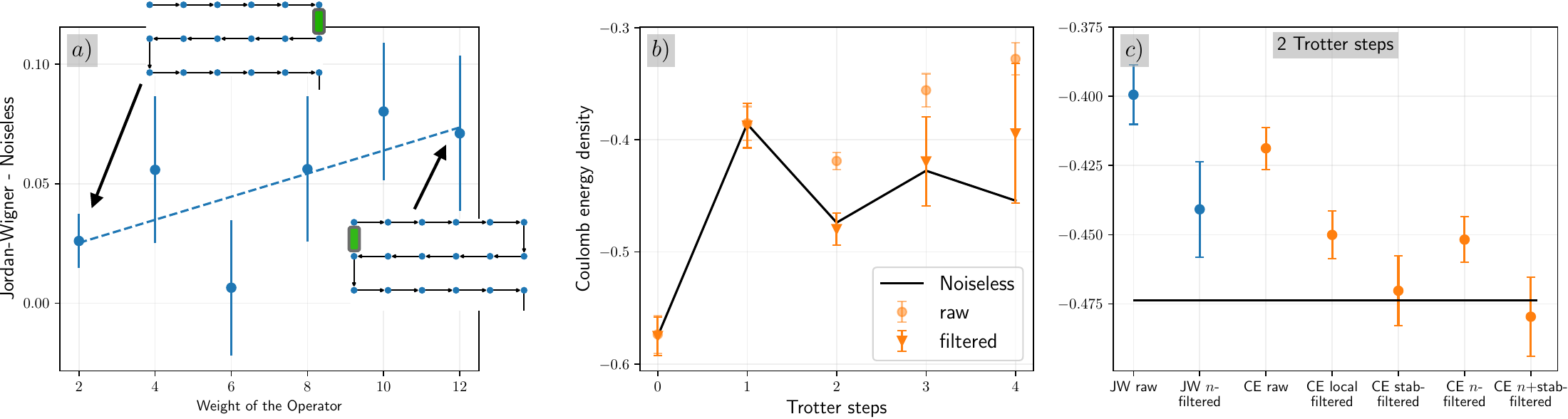}
\caption{\textbf{Sensitivity to noise in the Jordan-Wigner encoding and error mitigation.} a) Expectation value errors of operators as a function of the weight assigned to them by the JW encoding, after two Trotter steps on the $6 \times 6$ lattice,
with two representative operators overlaid. In contrast, all hopping operators in the bulk of the the Compact Encoding have weight 3. The dashed lines represent a fit $a \, \mathrm{weight} + b$ taking into account experimental uncertainties. The slope is $a = 0.0048 \pm 0.0025$. b) Effect of global filtering error mitigation on the Coulomb energy $e_\mathrm{INT}$ for the $6\times 6$ Compact Encoding. For this technique, only those shots are taken into account where the particle number $n=L^2/2$ and all measured stabilisers are equal to $+1$. c) Effect of different filtering techniques at 2 Trotter steps for the $6 \times 6$ lattice. For clarity, 200 shots were taken for JW and 400 shots were taken for CE.
\label{fig_weights}}
\end{figure*}

Is this scaling a disadvantage of JW due to a large gate count at $L=6$ (the susceptibility problem) or because of the extensive-weight observables appearing in the hopping energy (the sensitivity problem)? We investigate the effect of noise on operators of different weight after two Trotter steps in Fig.~\ref{fig_weights}. Indeed, we observe a trend for higher-weight operators to be more affected by noise, although the magnitude of this effect appears to be smaller than the extra noise introduced by the larger number of two-qubit gates, (720 for JW vs. 516 for CE, cf. Fig.~\ref{fig_gatecounts}b).

What is the source of the remaining system-size dependence of the loss of signal for the CE? In section \ref{sec_stabiliser_data_implies_memory_error}, we use data on the stabilisers of the compact encoding to reject the hypothesis that this scaling comes exclusively from extensive sensitivity ($\chi$) and show that it must instead be caused by extensively scaling hardware errors per gate ($\varepsilon \propto L$). This is compatible with the fact that larger circuits take longer to run even at the same circuit depth, which allows more memory error to accumulate from idling (roughly doubling between $L=4$ and $L=6$ \cite{decross_computational_2024}).

The physics of the states thus generated is shown in Fig.~\ref{fig_physics}. In Fig.~\ref{fig_physics}a and b we show the location of the $T=2$ state in the many body spectrum of the half-filled sector on the $4 \times 4$ lattice, for which full diagonalisation is achievable. As the mean-field initial state is already close to the edge of the spectrum, it is expected that virtually all hardware errors will drive the system to a higher-energy state. Fig.~\ref{fig_physics}c (d) shows the values of density-density correlations for the JW (CE) encoding. At half-filling on a bipartite lattice, one expects staggered charge ordering to occur at low effective temperatures and strong repulsive interactions. Such correlations do indeed develop for both encodings. What's more, unlike the energy, which acquires a minimum for two Trotter steps, charge density correlations do not saturate there, but keep increasing, with the strongest ordering appearing for the deepest circuits we ran. We conclude that the density matrices prepared in the experiment are different from thermal states corresponding to the given energies, whose correlation functions would become less pronounced at higher effective temperatures.

\section{Error Mitigation}
To learn more about the noisy density matrix prepared in these experiments, we investigate the effectiveness of different error mitigation techniques, focusing on the Coulomb energy density
\begin{equation}
\begin{aligned}
    e_\mathrm{INT} = \frac{ \braket{\psi |U^\dagger \left(V \sum_{\braket{ij}} n_i n_j \right) U| \psi}}{2L(L-1)}.
\end{aligned}
\end{equation}
A common technique to mitigate errors in a system with conserved quantities is to postselect shots that are compatible with expectation. The adiabatic evolution~\eqref{eq_adiabatic_evolution} preserves the overall particle number $n$. Additionally, for the compact encoding, we have measured simultaneously a subset of stabilisers $A_s$ in each shot, cf. Fig. \ref{fig_measurement_bases}a,b. We refer to the postselection of shots which fulfil $n=L^2/2$ and $A_s = +1$ as ``global filtering", and show the results of this technique in Fig.~\ref{fig_weights}b.




Global filtering shows impressive performance for the system size considered here, especially for the circuits with 2 Trotter steps. Ultimately, though, we expect it to scale more and more poorly with increasing number of qubits. Roughly speaking, this is because the probability of no error occurring decreases as $\exp(-\varepsilon T N)$ (for a quantitative overview of this issue, see section~\ref{sec_non_scalability_particle_filtering}). Thus, global filtering is an acceptable strategy only in the regime where the total number of errors is $\lessapprox 1$.

On the other hand, we have observed in Fig.~\ref{fig_main_result} that the loss of signal on local observables is much milder. This implies that shots discarded by global filtering contain valuable information: In the limit of large system size, an error introducing e.g., a single spurious fermion would leave intensive observables like the energy density largely unaffected, and it is wasteful to discard the shot. Is there any way to instead systematically use this information?

Indeed, suppose that an $X$ or $Y$ error occurred just before the final measurement. This will flip two of the stabilisers, which upon measurement will return $-1$. For lattices with $L>4$ and periodic boundary conditions, the location of the error is uniquely pinpointed by the stabiliser measurement syndrome~\cite{bausch_mitigating_2020}. Since the location of the error is known, the shot does not have to be entirely discarded. The bits that are unaffected by the error can be used for computations of observables located far away from the error, which reduces the information discard rate.

In practice, the situation is more complicated since the errors can occur at any point during the circuit and thus the error will gradually propagate through the system with some velocity $v$, while the stabiliser syndrome will remain frozen in time cf. Fig.~\ref{fig_errors}. Thus, to apply the local filtering error mitigation strategy, one will have to discard all bits within a chosen radius $r$ from the faulty stabiliser. This radius is ideally chosen to be the butterfly velocity of the error in question~\cite{kechedzhi_effective_2024}, but simply choosing the circuit light cone will already give an improvement over global filtering for shallow enough circuits. If the timestep when the stabiliser flipped from $+1$ to $-1$ is known, then $r$ can be set to the distance over which the error could propagate between the occurrence of the error and the final measurement. As the number of steps $T$ increases, $r$ becomes of the order of the size of the system and local filtering becomes equivalent to global filtering. However, for relatively shallow circuits such as the ones used in the current work, we show in Fig.~\ref{fig_weights}c that this ``local filtering" can significantly reduce the shot overhead to obtain an estimate of local observable to a given precision. Generally, local filtering can be used in a larger regime than global filtering, namely one where the union of the light-cones associated with all errors is smaller than the system size, i.e. $\varepsilon (v T)^d \lessapprox 1$ for a system of dimension $d$.

Finally, for deeper circuits, knowledge of the correlations between stabiliser syndromes and the observables of interest allows one to do Zero Noise Extrapolation without having to run the circuit at different noise levels. We give an overview of this method as well as a test case in section~\ref{sec_stabiliser_ZNE}.

\section{Conclusion}
We have experimentally realised the scaling advantage of the Compact over the Jordan-Wigner fermionic encoding. To achieve this, we have introduced a ``corner-hopping" scheme to reduce the number of gates required to simulate the hopping terms in the qubit Hamiltonian and applied this strategy to circuits encoding an adiabatic evolution towards the ground state of a spinless Hubbard model with up to $6 \times 6$ orbitals. We found the loss of signal in this experiment to depend only weakly on system size. Extrapolating Fig.~\ref{fig_main_result}c to $L^2=100$ modes using a signal-to-noise ratio scaling as $\exp(-\alpha L T)$, we expect to retain a signal of roughly $20\%$ for 20 Trotter steps. This signal may still be recoverable by error mitigation and falls into a regime that is challenging for classical methods, especially if a large part of the circuit is used for strongly entangling real-time dynamics. Since a large part of the observed system-size dependence is likely due to memory errors, we expect that the use of a suitable dynamical decoupling scheme would further reduce this dependence and improve the signal. On larger systems, memory errors coming from initial state preparation can also be suppressed by preparing the initial toric code state in constant depth using measurement and feed-forward~\cite{iqbal_topological_2024, foss-feig_experimental_2023}. All of these findings highlight the crucial importance of locality in quantum simulation and should motivate further work to carry over its benefits to other settings, including quantum chemistry. They also suggest that classically challenging Fermi-Hubbard simulations can be done without error correction on digital quantum computers---as long as gate and memory errors deteriorate only mildly when scaling up current systems.

For the model considered in this work scenario, a standard adiabatic algorithm was found to work exceptionally well at small number of Trotter steps due to ``Floquet Cooling" (cf. section~\ref{sec_trotter_error_encodings}),
Whether this phenomenon can be exploited more generally, and combined with other opportunities for speeding up adiabatic algorithms ~\cite{xie_variational_2022, kovalsky_self-healing_2023, tang_exploring_2024,schiffer_adiabatic_2022} remains an important open problem.


We have focused in this work on a spinless model to maximise the size of the available lattices. In practice, the spin degrees of freedom are essential for much of the interesting physics of the Hubbard model. The spinful model can be implemented by taking two layers of the model used here and we expect break-even to occur at a similar lattice size, which requires twice the number of qubits used here. In a model with on-site rather than nearest-neighbour interactions, the total gate cost is dominated even more by the hopping terms and so the advantage of the corner-hopping scheme over the standard decomposition will be larger (cf. section~\ref{sec_gate_count_and_weight}).

The question of break-even is more subtle for more realistic models that contain multiple species of fermions~\cite{clinton_towards_2024}, as well as $SU(2)$-breaking spin-orbit terms. In all of these models, the error detection properties of a given local encoding will play a crucial role. While higher-distance encodings have been developed~\cite{setia_superfast_2019} and tested~\cite{chien_simulating_2023}, there is a fundamental tradeoff between the ability to detect errors and easy implementation: errors that are identical to terms in the Hamiltonian can never be distinguished from our intentional application of those terms and so if the weight of those terms is small (and thus easy to implement), a larger number errors will go undetected. A related theoretical issue is whether we can learn something general about ``Hamiltonian errors", i.e. errors that consists of applying terms of the simulated Hamiltonian randomly in time and space.

\bibliography{references}

\section*{Acknowledgements}
This work was made possible by a large group of people, and the authors would like to thank the entire Quantinuum team for their many contributions. We are grateful for helpful discussions with and feedback from Mohsin Iqbal, Gabriel Greene-Diniz and Sheng-Hsuan Lin. E.G. acknowledges support by the Bavarian Ministry of Economic Affairs, Regional Development and Energy (StMWi) under project Bench-QC (DIK0425/01). K.H. and H.D. acknowledge support by the German Federal Ministry of Education and Research (BMBF) through the project
EQUAHUMO (grant number 13N16069) within the funding program quantum technologies - from basic research to market. The bulk of the experimental data reported in this work, including all circuits described in Fig.~\ref{fig_main_result}, was produced by the Quantinuum H2-1 trapped ion quantum computer, Powered by Honeywell, between 10-14th of June 2024.

\section*{Data availability}
The numerical data that support the findings of this study, including a full list of shots is available on the Zenodo repository~\cite{nigmatullin_supporting_2024}.

\section*{Author contributions}
R.N. designed the adiabatic path, generated all circuits, did a large part of the data analysis and invented the local stabiliser filtering technique. K.H. did many of the classical statevector and matrix product state calculations. S.M., D.G., P.S., M.M., T.G and N.H built the experiment and took the data. K.G., R.N. and K.H. made a library that was heavily used to automatically visualise and create circuits from fermionic encodings specified by edge and vertex operators. E.G. contributed to the theory of the Stabiliser-based Zero Noise Extrapolation as well as the visualisation. R.N., K.H., K.G, E.G. and H.D. conceived the experimental design. H.D. contributed to the compilation scheme, data analysis, initial hypothesis and drafted the initial manuscript, to which all authors contributed.



\clearpage
\onecolumngrid
\section*{Supplementary information}

\subsection{Implementing the hopping interaction in the Jordan-Wigner encoding using fermionic SWAP networks}
\label{sec_fSWAP_networks}

The JW encoding for a 2D rectangular fermionic lattice maps each site to a qubit ordered along a 1d line. The hopping interactions $c^\dagger_i c_j$ are mapped to

\begin{equation}
    c^\dagger_i c_j+c^\dagger_j c_i = \frac{1}{2}(X_i X_j + Y_i Y_j)Z_{i+1} ... Z_{j-1} 
\end{equation}
The string of the Pauli-Z operators acting on the sites between $i$ and $j$ is known as the JW string. The hopping terms between the non-JW adjacent sites can contain long JW strings strings - of the order of $L$ for $L\times L$ lattice. While these high weight Pauli operators can be decomposed into two-qubit gates using Pauli gadgets~\cite{cowtan_phase_2020}, a more efficient compilation strategy is to use fermionic SWAP (FSWAP) networks \cite{kivlichan_quantum_2018,cade_strategies_2020}. The idea of FSWAP networks is to move the qubits that are in non-JW adjacent positions into the adjacent positions using a sequence of FSWAP gates and then implement the local hopping interaction. The FSWAP gate is given by $\textrm{FSWAP} = \textrm{SWAP} \cdot \textrm{CZ}$, where SWAP is the two qubit swap gate and CZ is the controlled-Z gate. It ensures that the correct parity on all qubits is always maintained. 

\begin{figure}[!ht]
    \centering    \includegraphics[width=0.5\linewidth]{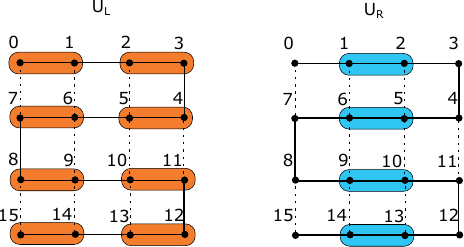}
    \caption{Fermi-swap operators $U_L$ and $U_R$, which are used to implement vertical hopping terms in the Trotterized dynamics in the JW encoding on a $4\times 4$ lattice.  $U_L$ swap odd numbered columns with those on their right and $U_R$ swap even numbered columns with those on their right. The snake JW ordering is used indicated by solid edges between the JW adjacent sites. The fermionic swaps between adjacent columns bring vertical hopping terms into JW adjacent position, between sites (3,4), (7,8) and (11,12), where they can be implement with 2 two qubit gates. All vertical hopping terms are implemented with $L_x$ rounds of $U_L U_R$, where $L_x$ is the number of columns. }
    \label{fig:fswap_network}
\end{figure}

In \cite{cade_strategies_2020}, an efficient FSWAP network for rectangular lattices was introduced and used to construct number preserving VQE ans\"{a}tze. Here, we adapt this FSWAP network for constructing the hopping layer in the Trotter dynamics circuits. 

For a rectangular lattice with $L_y$ rows and $L_x$ columns, the snake JW ordering is used such that all of the horizontal bonds are JW adjacent. Figure \ref{fig:fswap_network} illustrates the set-up on the $4\times 4$ lattice. First, the horizontal hopping terms are implemented, which are given by $e^{i\theta X_i X_{i+1}} e^{i \theta Y_i Y_{i+1}}$ with $\theta = -\tau t/2$. Up to single-qubit gates, these are equivalent to ZZphase, $e^{i \theta Z_i Z_{i+1}}$, gates which are natively available on Quantinuum H2 device. Odd horizontal bonds are implemented first, followed by the even bonds. Next, the vertical hopping terms are implemented using the FSWAP network. Let $U_L$ and $U_R$ be the operators that apply the FSWAP gates between the adjacent columns. $U_L$ swaps odd numbered columns with their neighbors to the right and  $U_R$ swaps even numbered columns with their neighbors to the right, as shown in figure \ref{fig:fswap_network}. An application of $U_L U_R$ brings a non-JW adjacent qubits into JW adjacent positions, which for the $4\times 4$ lattice are edges (3,4), (7,8) and (11,12). Once the qubits are in the JW adjacent positions, the hopping terms are applied in the same way as for the horizontal bonds. After $L_x$ applications of $U_L U_R$ all of the non-JW adjacent vertical hopping terms are implemented and the original qubit order is restored. On the architectures with all-to-all qubit connectivity such as the Quantinuum H2 device, the FSWAP gates are realized with only one two-qubit gate, CZ, since the swap operation can be implemented physically moving and swapping the qubit ions.

\subsection{State Preparation and Measurement}
\label{sec_state_prep_and_measurement}
\begin{figure*}[!ht]
	\centering
\includegraphics[width=1.0\textwidth]{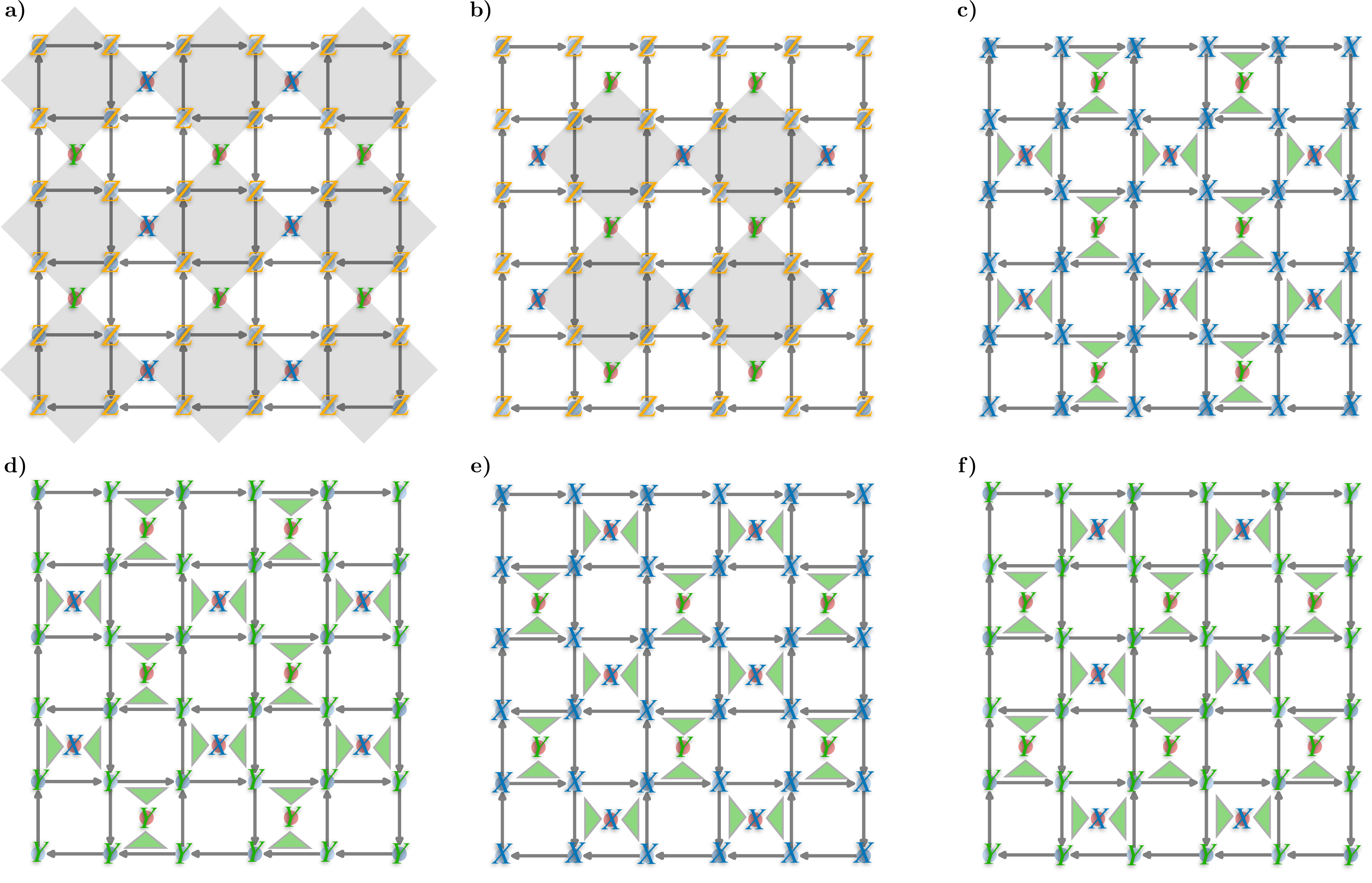}
\caption{\textbf{Measurement in the Compact Encoding.} We measure both kinetic and potential energy terms in the Compact Encoding of~\eqref{eq_hamiltonian} using 1-local measurements without any extra circuitry. a,b) Local bases to measure the interaction energy $e_\mathrm{INT}$. Both settings contain information about all bonds but allow for the simultaneous measurement of different sets of stabilisers (shaded gray). Additional stabilisers could be measured simultaneously by adding an ancilla and measuring $X\otimes Y \otimes X \otimes Y$ using 4 extra two-qubit gates for each stabiliser. c-f)  Local bases to measure the hopping energy $e_\mathrm{HOP}$. In principle, since particle number is conserved, measuring e.g., (c) and (e) would be sufficient. Additional stabilisers could be measured simultaneously by adding an ancilla and measuring $Z^{\otimes 4} \otimes X\otimes Y \otimes X \otimes Y$ using 8 extra two-qubit gates for each stabiliser.\label{fig_measurement_bases}}
\end{figure*}

To measure the energy density of interaction ($e_\mathrm{INT}$) and hopping energy ($e_\mathrm{HOP}$) in both Jordan-Wigner and Compact encoding, one must choose a suitable grouping of terms into commuting sets as well as a measurement strategy.  For the Compact Encoding, after executing state preparation and adiabatic time evolution, $e_\mathrm{INT}$ ($e_\mathrm{HOP}$) is obtained by collapsing the qubits locally in the bases shown in Fig.~\ref{fig_measurement_bases}a,b (c-f). The measurement strategy for the Jordan-Wigner encoding is described in the main text, section~\ref{sec_model}.

\begin{figure*}[!ht]
	\centering
\includegraphics[width=1.0\textwidth]{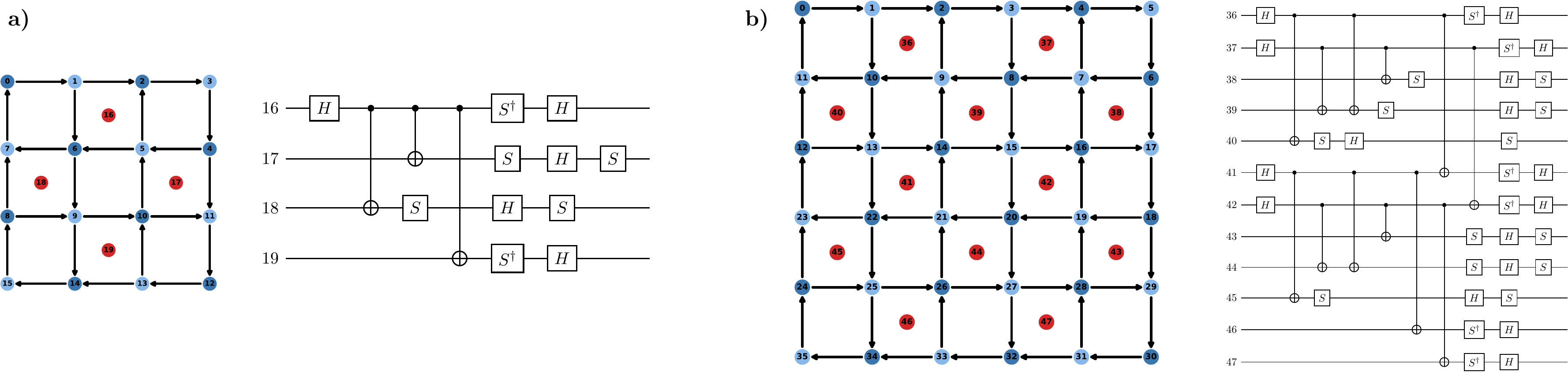}
\caption{\textbf{Initial state preparation for the Compact Encoding.} The fermionic exchange statistic in the compact encoding is realised locally by initialising the ancillae (red) in the ground state of a toric code. a) Circuit to prepare the toric code on the $4 \times 4$ lattice, using 3 two-qubit gates  b) Circuit to prepare the toric code on the $6 \times 6$ lattice, using 12 two-qubit gates \label{fig_toric_code_prep}}
\end{figure*}

While preparation of the Neel product state is trivial for the Jordan-Wigner encoding, in the compact encoding, one needs to initialise the system in the ground state of a toric code, such that the stabilisers shown in Fig.~\ref{fig2} and~\ref{fig_measurement_bases} are all equal to +1. To do this, we use the circuits shown in Fig.~\ref{fig_toric_code_prep}. The inital state we consider is a local eigenstate of $Z$ on the vertex qubits, and thus the toric code preparation acts only on the ancillae. For larger systems, it may be advantageous to prepare the toric code with measurement and feed-forward using a constant depth protocol~\cite{iqbal_topological_2024,foss-feig_experimental_2023}. For the system size we consider, we found the unitary preparation to require only minimal resources and work well in practice.

Going forward, we suspect that it will be advantageous to add ancillae on the outside of the system, even when simulating open boundary conditions. Access to the maximal number of stabilisers ($N/2$) allows for more powerful error mitigation. In this case, when measuring diagonal operators like density correlations, one can add $N/4$ ancillae (or, in the other extreme, use one ancilla repeatedly) and measure coherently the off-diagonal part ($X \otimes Y \otimes Y \otimes X$) of the stabilisers that are not measured destructively (cf. Fig.~\ref{fig_measurement_bases}a). This can achieved with $4 \times N/4 = N$ extra two-qubit gates. For measuring hopping operators, one can simultaneously obtain all stabilisers by using $8 \times N/4 = 2N$ two-qubit gates to coherently measure the stabilisers whose off-diagonal part is not measured destructively (e.g., for the setting in Fig.~\ref{fig_measurement_bases}c, these would be the stabilisers shown in Fig.~\ref{fig_measurement_bases}a) and then use $4 \times N/4 = N$ two-qubit gates to measure $Z^{\otimes 4}$ of the complementary set of stabilisers---for a total of $3N$ extra two-qubit gates.

Independent of the final measurement, it may be interesting to measure all stabilisers coherently in the middle of the circuit, as this would allow for limited active error correction and strengthen the local error mitigation technique developed in the main text. In fact, currently existing devices do have the capability to execute many of these mid-circuit measurements in a shot with high fidelity~\cite{iqbal_topological_2024, iqbal_non-abelian_2024, granet_noise-limiting_2024}. The full coherent measurement of all stabilisers requires $8 \times N/2 = 4N$ two-qubit gates. These extra costs are smaller than those associated with implementing a single Trotter step, which would be $7N$ for a layer of fermionic hopping using the corner hopping scheme, and $2N$ for a layer of interaction ($N/2$ for a spinful Hubbard model with a total of $2 \times N/2$ modes and on-site interactions only).

In principle, rather than localising the hopping and stabiliser operators on individual ancillae, one can attempt to simultaneously diagonalise them, i.e. apply some circuitry such that the full (commuting) set of desired operators is diagonal (but not necessarily local) in the $Z$-basis. Using the companion code of~\cite{berg_circuit_2020}, we have found that, to diagonalise a maximal subset of commuting hopping operators plus all stabilisers in our system would require 392 two-qubit gates for the $6 \times 6$ system and is thus not competitive for the final measurement for these observables.

\subsection{Fermionic edge and vertex operator mappings}

The fermion-to-qubit mapping expressed in terms of so-called vertex and edge operators defined on the connectivity graph between the fermionic modes. This formalism facilitates the construction of the encoded Hamiltonian. For completeness, we summarise the connection between the vertex and edge operators, $\{ V_i \}$ and $\{ E_{ij} \}$, and the fermion creation and annihilation operators $\{ c^\dagger_j,c_j \}$ (see, for example, \cite{derby_compact_2021}\cite{setia_superfast_2019}\cite{chien_simulating_2023} for more details).

The vertex and edge operators are signed Pauli strings, which satisfy the following relations:
\begin{align}
E_{jk} & = - E_{kj} \\
E_{jk} V_k & = - V_l E_{jk} \\
E_{jk} E_{kl}& = - E_{kl} E_{jk},
\end{align}
where $j$, $k$ and $l$ are distinct indices. 
In terms of edge and vertex operators, the quadratic fermion operators are given by

\begin{equation}
\begin{aligned}
    c^\dag_j c_k &= \frac{i}{4} (1-V_j)E_{jk} (1-V_k) \\
    c_j c_k^\dagger &= \frac{i}{4} (1-V_j)E_{jk} (1+V_k)\\
    c_j c_k &= \frac{i}{4} (1+V_j)E_{jk} (1-V_k)\\
    c_j^\dagger c_k^\dagger &= \frac{i}{4} (1+V_j)E_{jk} (1+V_k)
\end{aligned}
\label{eq:cdagj_ck_mapping}   
\end{equation}
Without loss of generality, one can set $E_{kk}=-i I$ such that

\begin{equation}
    c^\dag_k c_k = \frac{1}{2} (1-V_k) \label{eq:cdagk_ck_mapping}
\end{equation}
The product of any loop of edge operators is a stabiliser of the fermionic quantum code. The irreducible loops generate the set of stabiliser generators $\{S_j \}$.

For a compact encoding the vertex encoding are given by \cite{derby_compact_2021}

\begin{equation}
V_j = Z_j  \label{eq:CE_vert_ops}
\end{equation}
and the edge operators are given

\begin{align}
      E_{ij} = \begin{cases}
                X_i Y_j X_a \quad (i,j) \textrm{ oriented downwards}\\
               -X_i Y_j X_a\quad (i,j) \textrm{ oriented upwards}\\
               X_i Y_j Y_a\quad (i,j) \textrm{ horizontal}
            \end{cases} \label{eq:CE_edge_ops}
\end{align}
where the index $a$ denotes the ancilla next to the edge $(i,j)$ and the orientation of the edges is shown in figure \ref{fig2}. 
One can verify that the mapped Hamiltonian terms (\ref{eq:CE_mapped}) can be obtained by 
substituting (\ref{eq:CE_vert_ops})-(\ref{eq:CE_edge_ops}) in (\ref{eq:cdagj_ck_mapping})-(\ref{eq:cdagk_ck_mapping}). For completeness and easy reference, the relevant expressions for nearest neighbour fermionic bilinears, assuming an arrow is pointing from $j \rightarrow k$, are
\begin{equation}
\begin{aligned}
c_j^\dagger c_k = \frac{iP_a}{4}\big[ X_j Y_k - i Y_j Y_k - i X_j X_k + Y_j X_k\big] \\
c_j c_k^\dagger = \frac{iP_a}{4}\big[ X_j Y_k + i Y_j Y_k + i X_j X_k + Y_j X_k\big] \\
c_j c_k = \frac{iP_a}{4}\big[ X_j Y_k + i Y_j Y_k - i X_j X_k - Y_j X_k\big] \\
c_j^\dagger c_k^\dagger = \frac{iP_a}{4}\big[ X_j Y_k - i Y_j Y_k + i X_j X_k - Y_j X_k\big] \\
\end{aligned}
\label{eq_bilinears}
\end{equation}
where $P_a$ is the Pauli operator $Y_a$, $X_a$ or $-X_a$ according to~\eqref{eq:CE_edge_ops}. The individual components of the corner hopping scheme can be read off by inverting~\eqref{eq_bilinears}:
\begin{equation}
\begin{aligned}
X_j X_k P_a &= c_j^\dagger c_k - c_j c_k^\dagger + c_j c_k - c_j^\dagger c_k^\dagger \\
Y_j Y_k P_a &= c_j^\dagger c_k - c_j c_k^\dagger - c_j c_k + c_j^\dagger c_k^\dagger.
\end{aligned}
\label{eq_XXP}
\end{equation}

The encoded states must in the +1 eigenspace of the stabilisers $\{S_j \}$. The initial states used in this work are product states, with each vertex either empty or occupied. Such product states are stabiliser states of the vertex operators $\{ \pm V_j \}$, where the signs are chosen to indicate presence or absence of particles. 

One advantage of edge and vertex operator formalism is that for any product initial state, one can always use the all-zero $|0..00\rangle$ state on the vertices and a toric state on the ancilla, with the presence of fermions on a given site $j$ indicated by redefining the vertex operator $Z_j \rightarrow - Z_j$. After redefining the vertex operators to fix the correct initial state, the Hamiltonian can then be constructed using mappings (\ref{eq:cdagj_ck_mapping}) and (\ref{eq:cdagk_ck_mapping}) and then Trotterized as described in the main text. This procedure ensures that the angles of the ZZphase gates in the Trotter circuits, $\theta_j$ (figure \ref{fig2}), have the correct sign. Another practical advantage of using the edge and vertex operators is the ability to easily switch between different encodings by redefining the edge and vertex operators.

\subsection{Generalised Corner Hopping}

\begin{figure*}[!ht]
	\centering
\includegraphics[width=1.0\textwidth]{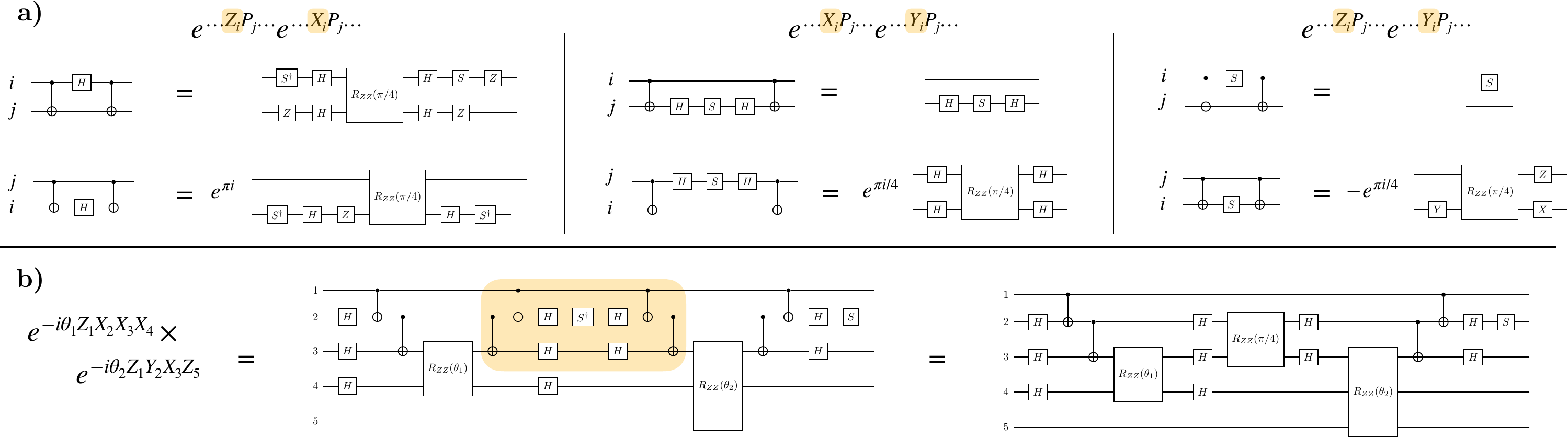}
\caption{\textbf{Generalised Corner Hopping.} (a) Elementary circuit identities that can be used repeatedly to remove gates from a circuit implementing Trotterised dynamics of multi-body operators. (b) Example of how the circuit identities can be used for a pair of adjacent $n=4$-body terms which share 3 qubits.   \label{fig_generalised_corner_hopping}}
\end{figure*}

The corner hopping scheme developed in Fig.~\ref{fig2} was essential for the compact encoding to be experimentally competitive with Jordan-Wigner. The scheme relies on judiciously ordering the 3-body terms in the Trotter expansion of $U_\mathrm{hop}$ and taking advantage of cancellations of type shown in Fig.~\ref{fig2}.
A natural question is whether the scheme can be generalised to (i) $n$-body operators with $n>3$ and (ii) basis changes different from $X \leftrightarrow Y$. This is indeed the case, as long as two sequential $n$-body operators share at least two qubits, and the basis ($X$,$Y$ or $Z$) is the same on at least one of the two qubits. In Fig.~\ref{fig_generalised_corner_hopping}a, we show all possible subcircuits that occur in this scenario (up to complex conjugation). The largest savings occur when changing $X \leftrightarrow Y$ or $Z \leftrightarrow Y$. When the basis remains the same on both qubits, the CNOT gates involved trivially cancel.

When two adjacent terms share more than two qubits, larger cancellations occur. For example, in Fig.~\ref{fig_generalised_corner_hopping}b, we show a product of two $n=4$-body terms which overlap on $k=3$ qubits, and only one of the qubits undergoes a basis change between the two terms. In this case, one can apply the equations in Fig.~\ref{fig_generalised_corner_hopping}a repeatedly to cancel three maximal entanglers. This strategy will likely be essential when implementing higher weight fermionic encodings whose larger code distance may provide some advantages regarding error mitigation and correction~\cite{setia_superfast_2019}.

How should one go about finding the ordering of terms which allows the maximum number of cancellations? Whenever there is a local lattice geometry, there are $\mathcal{O}(1)$ candidate orderings, independent of system size, because distant terms do not share qubits. In that case, one may consider a single local arrangement (for example the triangles considered in the main text of this work) and brute force search through all $\mathcal{O}(1)$ permutations, applying Fig.~\ref{fig_generalised_corner_hopping}a repeatedly to the circuit corresponding to each permutation.

The corner-hopping scheme can also  be used to reduce fermionic simulation costs on other lattices, for example by applying it to two thirds of the bonds of a honeycomb lattice and executing the remaining bonds in the standard way using Pauli gadgets. Similarly, the same identities as shown in Fig.~\ref{fig2} can directly be applied to higher-weight operators as they appear in other fermionic encodings~\cite{setia_superfast_2019}. There may also be other applications outside of fermionic encodings where similar sequences of higher-weight operators appear.

\subsection{System-Size dependent noise in the Compact Encoding cannot be due to sensitivity $\chi$ alone}
\label{sec_stabiliser_data_implies_memory_error}
\begin{figure*}[!ht]
	\centering
\includegraphics[width=1.0\textwidth]{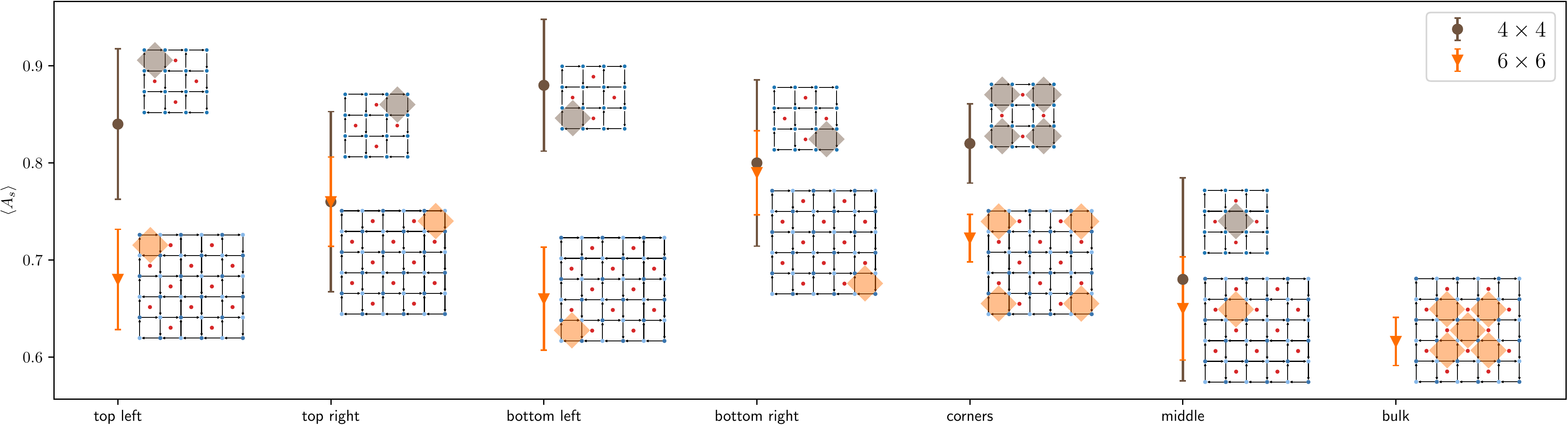}
\caption{\textbf{Noise on the stabilisers in the compact encoding.} Stabilisers on the $6 \times 6$ system are significantly more affected by noise than on the $4 \times 4$ system (noiseless expectation value is $\braket{A_s} = +1$). As shown in Fig.~\ref{fig_errors}a, this is incompatible with a model where the error per gate is constant with system size $\varepsilon = \mathcal{O}(1)$ and must therefore be caused by memory errors in the device. Corner stabilisers have Pauli weight 6 and middle/bulk stabilisers have Pauli weight 8. ``Corners" and ``Bulk" refer to averages over all stabilisers shown in the pictograms. \label{fig_stabilisers4vs6}}
\end{figure*}
In the main text, we found the impact of noise to scale signicantly more favourably in the compact vs. the Jordan-Wigner encoding. However, our data in Fig.~\ref{fig_main_result}d suggests that, even for the compact encoding, larger system sizes are associated with greater loss of signal at the same number of Trotter steps. In the language of equation~\eqref{eq_snr}, we would like to study whether this behaviour is caused by (i) extensive sensitivity of local observables to noise ($\chi \propto L)$ or (ii) errors in the hardware that increase the effective two-qubit error \textit{per gate} when increasing the system size ($\varepsilon \propto L$). Understanding this issue is essential: Case (i) implies a fundamental issue simulating dynamics of the type considered in this work, independent of the digital quantum computing platform used, and may well imply that quantum error correction is required for simulating large-scale systems. Case (ii), in contrast, can be managed by identifying the responsible sources of error in the particular quantum computer that is used and attempting to address them either in hardware or software.

We can resolve this question to a large extent by considering the noisy data not for the energy density but instead for the expectation values of the stabilisers $A_s$ in the compact encoding (the operators are shown in shown in Fig.~\ref{fig_measurement_bases}a,b). Crucially, as shown in Fig.~\ref{fig_errors}a, each of the stabilisers commutes with all components of the circuit, i.e. each corner hopping and each interaction component individually. This means that errors on qubits that are not acted on by a given stabiliser $A_s$ do not affect the measured value of that stabiliser (which is different from a light-cone argument as the stabiliser does not grow in time). This implies that, for a given number of Trotter steps $T$, the stabilisers see exactly the same (number of) noisy components independent of system size. If the failure probability of these components was independent of system size (i.e. $\varepsilon = \mathcal{O}(1)$) then the expectation value of $A_s$ on different system sizes must be equal.

Figure~\ref{fig_stabilisers4vs6} shows that this is not the case. For example, the mean expectation value of the corner stabilisers after $T=2$ Trotter steps at $L=4$ ($L=6$) is $0.82 \pm 0.04$ ($0.72 \pm 0.02$) which implies a probability of roughly $98\%$ that the corner stabilisers on the $4 \times 4$ system are less noisy, and therefore the hypothesis that $\varepsilon = \mathcal{O}(1)$ can be rejected. Indeed, this finding is compatible with Fig A3c in \cite{decross_computational_2024} which shows that the amount of memory error per gate on the device roughly doubles between the system sizes corresponding to the $L=4$ and $L=6$ compact encodings in this work. Intuitively speaking, this is due to the constant number of gate zones (4) on the devices which means that more ion swaps must be carried out to execute one layer of $N/2$ gates.

How much of the system-size dependence on the energy density shown in Fig.~\ref{fig_main_result}d is due to this memory error? While we cannot know the errors that occurred during the experiments with perfect precision, we can compare the scaling of the noise on the energy density data with the scaling of the noise on the stabilisers (which we know must be due to memory errors). Using as $A_s$ the average of the corner stabilisers, we find that
\begin{equation}
\begin{aligned}
    \frac{\braket{A_s^\mathrm{noisy}(L=6)}}{\braket{A_s^\mathrm{noiseless}(L=6)}} &= \left(\frac{\braket{A_s^\mathrm{noisy}(L=4)}}{\braket{A_s^\mathrm{noiseless}(L=4)}}\right)^{1.64 \pm 0.44}, \quad \quad \mathrm{while} \\
    \frac{\braket{H^\mathrm{noisy}(L=6)}}{\braket{H^\mathrm{noiseless}(L=6)}} &= \left(\frac{\braket{H^\mathrm{noisy}(L=4)}}{\braket{H^\mathrm{noiseless}(L=4)}}\right)^{1.65 \pm 0.75}.
\end{aligned}
\end{equation}
Therefore, it is possible that the entirety of the increased loss of signal on the energy density from $L=4$ to $L=6$ is due to memory error, i.e. the aforementioned case (ii). We conclude that, in order to simulate large-scale Fermi-Hubbard systems with the quantum charge-coupled device architecture used in this work, one should prioritise the minimisation of memory errors, either on the hardware level (e.g., by increasing the density of gate zones in the device) or on the software level (e.g., by using dynamical decoupling techniques).

\subsection{Implementation details for the Local Filtering technique}
The flavour of local filtering used to obtain the data in Fig.~\ref{fig_weights}c is as follows: In the experiments, only a partial set of stabilisers, was measured simultaneously with the interaction energy, cf. Fig~\ref{fig_measurement_bases}a,b. While the partial stabiliser information does not allow us to uniquely pinpoint the location of the error, a simplified local filtering procedure was applied, where any bits which overlap with the faulty stabilisers are discarded when computing the expectation of a local observable. As shown in Fig.~\ref{fig_weights}c, the local filtering error mitigation significantly improves the estimate of $e_\mathrm{INT}$, while only increasing the standard error by $13\%$ with respect to the raw data (compared to $88\%$ for global filtering). We find that on average, the shot discard rate when computing a local observable after local filtering is $18\%$.

\subsection{Stabiliser-based Zero Noise Extrapolation}
\begin{figure*}[!ht]
	\centering
\includegraphics[width=1.0\textwidth]{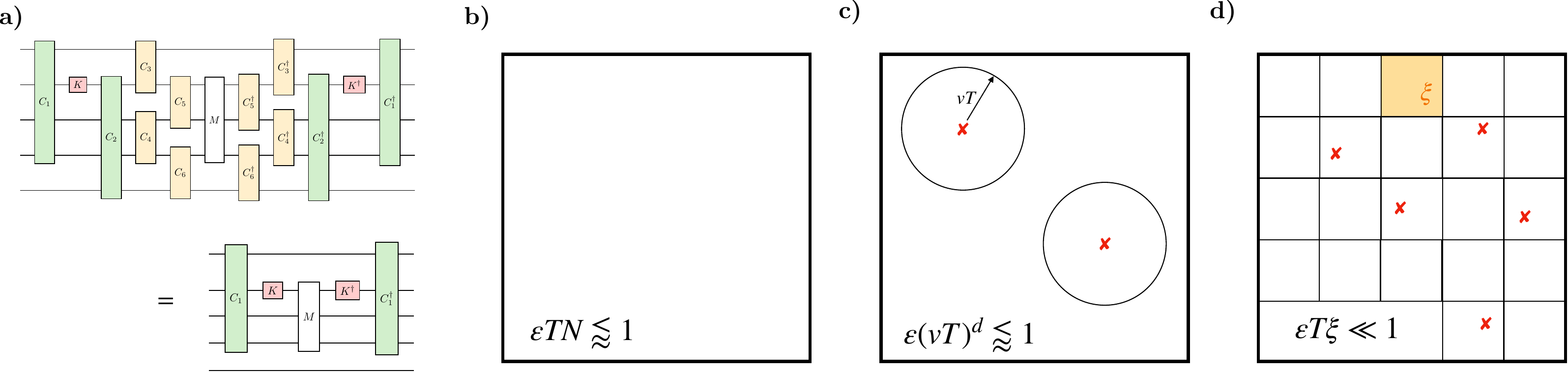}
\caption{\textbf{Effect of errors on stabilisers $M$ and three different error mitigation techniques.} a) One can group the circuit into components $C_i$ (orange and green) and work in a digital error model where each component is associated to a quantum channel $\mathcal{C}_i(\rho) = (1-\varepsilon_i) C_i \rho C_i^\dagger + \varepsilon_i \sum_j K_{ij} C_i \rho C_i^\dagger K_{ij}^\dagger$ with errors operators $K$ (red). For $M$ a conserved quantity, like particle number or the stabilisers of the compact encoding, errors look like they are frozen in time and they can never propagate to flip a stabiliser far away from where they occur. b) Global filtering works in the regime where the overall number of errors is small. c) Local filtering works in the regime where errors do not have time to propagate too far from the position at which they occur (and where they cause a stabiliser flip). d) Local observable extrapolation allows for a larger number of errors, only requiring multiple errors to not occur within a single stabiliser region $\xi$. \label{fig_errors}}
\end{figure*}

In the main text we have studied the efficacy of both a global and a local filtering technique that is based on postselecting shots with compatible stabiliser (and particle number) values. The global filtering is effective in the regime where the total number of errors is small $\varepsilon T N \lessapprox 1$ (Fig.~\ref{fig_errors}b). The local filtering technique can be applied for larger number of qubits, as long as the circuits are sufficiently shallow, i.e. $\varepsilon (v T)^d \lessapprox 1$ (Fig.~\ref{fig_errors}c). This limitation originates from the fact that errors look frozen in time for conserved quantities like the stabilisers (cf. Fig.~\ref{fig_errors}a), whereas errors do generally propagate to affect local observables of interest in a light cone that does grow with time. The purpose of this section is to outline a method with even milder conditions, requiring only $\varepsilon T \xi \ll 1$, i.e. the number of errors within the area of a stabiliser $\xi$ is much smaller than 1 (Fig.~\ref{fig_errors}d).

\label{sec_stabiliser_ZNE}
\subsubsection{The mitigation technique}
To start, consider grouping all gates in the circuit into $V=DN$ components $C_i$, as shown in Fig.~\ref{fig_errors}a. Here, $D$ is the average circuit depth which is linearly related to the number of time steps $T$ used in the main text and $V$ refers to the total circuit volume rather than the Coulomb interaction throughout this section. In abuse of notation, in the rest of this section we refer to $\varepsilon$ as the probability of error \textit{per gate}, which is different from (but again linearly related to) the use of $\varepsilon$ in the main text, where it denotes the probability an error occurring per qubit per time step. Each of the noisy components can be associated with a channel $\mathcal{C}_i(\rho) = (1-\varepsilon_i) C_i \rho C_i^\dagger + \varepsilon_i \sum_j K_{ij} C_i \rho C_i^\dagger K_{ij}^\dagger$. We can then expand the noisy expectation value of an observable $M$ as
\begin{equation}
\begin{aligned}
    M(\varepsilon) &= \sum_{k=0}^V (1-\varepsilon)^{V-k} \varepsilon^k \binom{V}{k} M_k
\end{aligned}
\end{equation}
where $M_k$ is the expectation value of $M$ in the sector with exactly $k$ errors, e.g.
$\binom{V}{1} M_1 = \sum_{ij} \mathrm{Tr} [M C_V \mydots C_{i+1} K_{ij} C_{i} \mydots C_1 \rho_0 C_1^\dagger \mydots C_i^\dagger K^\dagger_{ij} C_{i+1}^\dagger \mydots C_V^\dagger]$. Now this expansion can be rewritten in orders of $\varepsilon$
\begin{equation}
\begin{aligned}
    M(\varepsilon) &= \sum_{l=0}^V (-\varepsilon)^l \binom{V}{l} \Delta^l M
\end{aligned}
\end{equation}
where we have defined the $l$-th order forward finite difference $\Delta^l M = \sum_{k=0}^l (-1)^k \binom{l}{k} M_k$.
If, in addition to the raw data $M(\varepsilon)$, one has an estimate of the values $M_k$ as well as the probabilities $p_k$ for $k$ errors to occur, one can systematically remove terms of increasing order in $\varepsilon$ from the sum to be eventually left with $M_0$, the noiseless value. For example, using $p_k=\varepsilon^k (1-\varepsilon)^{V-k}{V\choose k}$, the estimator
\begin{equation}
\begin{aligned}
        \hat{M}^{(1)} &:= \sum_{k=0}^{V-1} p_k \left[ (1+\varepsilon V) M_k - \varepsilon V M_{k+1} \right] \\
        &=M_0 + \mathcal{O}(\varepsilon^2)
\end{aligned}
\end{equation}
 reduces the residual bias to order $\varepsilon^2$. If $\varepsilon$ is not known, one can estimate it from the probability of observing $k$ errors as
 \begin{equation}
     \varepsilon = \frac{p_{k+1} \binom{V}{k}}{p_k \binom{V}{k+1}} \left(1+\frac{p_{k+1} \binom{V}{k}}{p_k \binom{V}{k+1}}\right)^{-1}\,.
 \end{equation}
 Higher order estimators can be created in the same way. To obtain an uncertainty $\epsilon$ on the estimator of order $n$, $\mathcal{O}\left( (\varepsilon V)^{2n} /\epsilon^2 \right)$ shots are required.

The key question is then: how large do we need to choose $n$ (the order of the estimator) in order to reduce the bias to less than a given value $b$? This depends on the smoothness of $M(\varepsilon)$ which is deeply related to the physics we are trying to simulate: A bound for that to occur is
\begin{equation}
    n > e \varepsilon D N^\alpha + \log (\beta/b)
\end{equation}
where we have introduced a constant $\beta$ independent of all other parameters that is chosen such that $\frac{V!}{(V-l)!} |\Delta^l M| \leq \beta( D N^{\alpha})^l$ for all $l$. Consequently, the number of shots to reach a bias less than $b$ with error bar bounded by $\epsilon$ is $\mathcal{O}\left( 
(\varepsilon D N)^{(2e\varepsilon DN ^\alpha + 2\log \beta/b)} \right/\epsilon^2)$. For a fixed circuit depth, we see that the method crucially relies on the scaling of $\Delta^l M$ in $N$. If $\Delta^l M$ is constant in $N$, then $\alpha = 1$ and the method just outlined requires a number of shots that is exponential in system size, similar to global filtering techniques.

For the simulation of local observables and Hamiltonians, generically $\alpha < 1$ due to a ``Dilution of Errors"\cite{granet_dilution_2024}, and there is evidence that even $\alpha = 0$ in a variety of generic simulation tasks. How can this occur? Consider for example $\Delta^1 M = M_0 - M_1$. The term $M_1$ contains error events like the single spurious fermion mentioned before, which may cause local observables to, on average, change by an amount $|\Delta^1| = 1/N$. If we have generally that $M_k = (1-\frac{1}{N}) M_{k-1}$ then $\Delta^l M = \mathcal{O}(1/N^l)$ and $\alpha=0$ follows. In this case, for a fixed total error per qubit $\varepsilon T$, one can obtain an estimator with essentially arbitrarily small bias and error bars with a number of shots polynomial in system size. In practice, one will be limited to small $n$, although there are techniques that can take into account higher orders more efficiently, based on further assumptions (which state that the effect of multiple errors can be factorised). We also note that the exact scaling with $N$ may be better in practice, due to spatial self-averaging effects.

Errors in simulations using the Compact Encoding are particularly well suited to be mitigated by the programme described before: First, due to the locality of all components, we expect Dilution of Errors to occur. What's more, the fact that extensively many local conserved quantities are available makes it possible to (approximately) implement the scheme described before without precise characterisation of the hardware errors and without having to run different circuits. This is because, for each shot, one not only has access to the value of the target observable (e.g., in our case $e_\mathrm{INT})$, but also a syndrome of the form $\mathbf{s} = (n, A_1, \dots)$, where $n$ is the particle number and $A_1, \dots = \pm 1$ are the value of the stabilisers. The syndrome information offers limited capabilities for active error correction. This is because two errors related by terms in the Hamiltonian lead to equivalent syndromes (and we note that this a fundamental property of local fermionic encodings---they are constructed in a way that the Hamiltonian terms commute with the stabilisers). Nevertheless, one can use the syndrome information to estimate \emph{how many} errors have occurred. This is especially relevant in the regime where there is a finite density of errors $\varepsilon TN > 1$ but $\varepsilon T \xi < 1$, where $\xi$ is the radius of a stabiliser. Before showing this strategy in a toy model, we note that this simplification applies any time that extensively many local conserved quantities exist, like in other fermionic encodings \cite{setia_superfast_2019} or certain lattice gauge theory simulations. We also note that the presence of conserved quantities does not preclude one from additionally measure the effect of errors on the observables of interest, if a good device characterisation is available (and so even errors that are not detected by the stabilisers can be mitigated in this way).

\subsubsection{Toy model: Ising model with pair encoding \label{sec_pairising}}

\begin{figure*}[!ht]
	\centering
\includegraphics[width=0.32\textwidth]{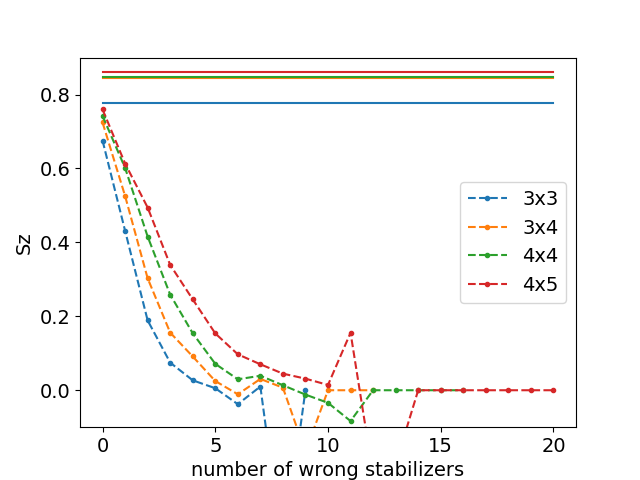}
\includegraphics[width=0.32\textwidth]{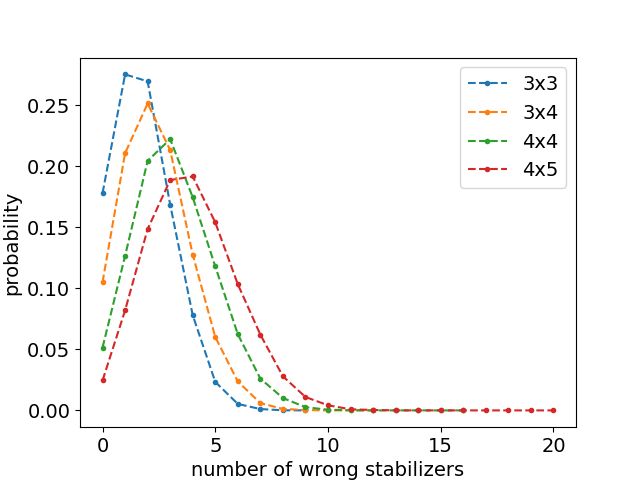}
\includegraphics[width=0.32\textwidth]{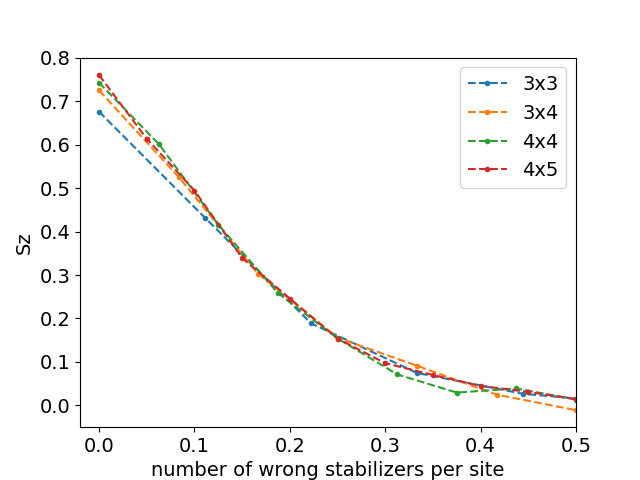}
\caption{\textbf{Stabilisers in the pair Ising model.} Left panel: expectation value of $S_z$ as a function of number of wrong stabilisers in the noisy simulation of Section \ref{sec_pairising}, with the solid lines indicating the noiseless expectation value, for different system sizes. Middle panel: probability of finding a given number of wrong stabilisers. Right panel: expectation value of $S_z$ as a function of number of wrong stabilisers divided by the number of sites.\label{fig_pairising}}
\end{figure*}

In order to be able to perform numerical simulations of the effect of noise on systems with a large number of stabilisers, we consider the following toy encoding. We define a ``logical" qubit as a pair of two ``physical" qubits, with the logical $|0\rangle$ being $|00\rangle$ and the logical $|1\rangle$ being $|11\rangle$. Given $L$ logical qubits on $N=2L$ qubits, there are thus $L$ stabilisers given by $Z_{2j}Z_{2j+1}$ for $j=0,...,L-1$. We implement a quantum quench in the 2D Ising model with this encoding. Namely, we define a unitary operator
\begin{equation}
    U=\exp\left(ih{\rm d}t\sum_{J=1}^L X^{\log}_J\right)\exp\left(i{\rm d}t\sum_{\langle J,K\rangle} Z^{\log}_JZ^{\log}_K\right)\,,
\end{equation}
for some parameters $h,{\rm d}t$, and where the exponent $\log$ indicates that the Pauli operators act on the logical qubits, and $\langle J,K\rangle$ that the logical qubits $J,K$ are neighbours on a 2D square lattice. In terms of the physical qubits, this operator is
\begin{equation}
    U=\exp\left(ih{\rm d}t\sum_{j=1}^L X_{2j}X_{2j+1}\right)\exp\left(i{\rm d}t\sum_{\langle j,k\rangle} Z_{2j}Z_{2k}\right)\,.
\end{equation}
We fix $h=1,{\rm d}t=0.1$ and wish to compute the expectation value of $S_z=\frac{1}{L}\sum_{J=1}^L Z_J^{\log}$ in the state $U^n|0\rangle^{\otimes L}$ for $n=100$. On a perfect, noiseless hardware, the measurement of each of the stabilisers $Z_{2j}Z_{2j+1}$ would systematically be $1$ at the end of the computation. We now model a noisy hardware by the application of a two-qubit depolarizing channel with amplitude $\lambda=0.001$ after every two-qubit gate $e^{i\theta XX}$ or $e^{i\theta ZZ}$. This realistic noise model does not commute with the stabilisers and one may measure $Z_{2j}Z_{2j+1}=-1$ for some stabilisers at the end of the noisy circuit. We present the numerical results in Figure \ref{fig_pairising} where $20000$ shots were simulated for different system sizes. In the middle panel, we observe as expected that the probability of finding a certain number $n$ of wrong stabilisers $Z_{2j}Z_{2j+1}=-1$ follows a Gaussian-like curve, with an average value that grows with system size $L$. In the left panel, we then plot the expectation value of $S_z$ restricted to the shots where $n$ wrong stabilisers are observed, and plot it as a function of $n$. Importantly, we observe that this expectation value behaves linearly with $n$ at small $n$. This allows one to perform a linear extrapolation to $n=0$ in this regime. We also observe that there is a discrepancy with the exact value even at $n=0$ wrong stabilisers: this comes from the fact that in this simple encoding, $Z$ errors are undetectable, and also from the fact that two $X$ errors occurring on the same logical qubit will be undetectable. In the right panel, we show the same expectation value of $S_z$, but as a function of $n/L$. We observe an approximate scaling collapse, indicating that single errors contribute only to $\mathcal{O}(1/L)$ to the expectation value, instead of order $\mathcal{O}(L^0)$ as a priori expected.

To evaluate the efficiency of the noise mitigation procedure we propose, we consider size $4\times 5$ with $1000$ shots, and perform a linear fit on the expectation values as a function of $n$ the number of wrong stabilisers, for $0\leq n\leq 5$, weighted by the error bars on each point. The average extrapolated value that we obtain at $n=0$ over different batches of $1000$ shots is $0.745$, compared to $0.759$ for the exact value, with a variance $0.00614$. If we postselected the shots with $n=0$, which corresponds in average to keeping only $\approx 25$ shots, the variance obtained would have been $3$ times larger for this specific case, requiring three times the number of shots to get to the same precision. This shows that this mitigation technique can harness information from shots with a non-zero  number of wrong stabilisers. At larger system sizes, the proportion of shots with $n=0$ wrong stabilisers would become exponentially small. This extrapolation method would be able to obtain an estimate of the expectation value at $n=0$ while using all noisy shots.

Finally, let us comment on the simulation technique we used. To be able to simulate in Fig \ref{fig_pairising} system sizes as large as $L=4\times 5=20$ pairs of qubits, so $40$ qubits, we used the following technique. To simulate depolarizing noise applied after every two-qubit gate, one can generate trajectories where Pauli strings among $IX$, $IY$, $IZ$, $XI$, $XX$, $XY$, $XZ$, $YI$, $YX$, $YY$, $YZ$, $ZI$, $ZX$, $ZY$, $ZZ$ are inserted randomly with equal probability after every two-qubit gate. Let us show that at any time point, the state of a trajectory lives in the vector space of dimension $2^L$ obtained by imposing that each pair of qubits is either in a linear combination of $|00\rangle$ and $|11\rangle$, or in a linear combination of $|01\rangle$ and $|10\rangle$, which we call property $\mathcal{P}$. Property $\mathcal{P}$ holds true at the initialization of the qubits. Property $\mathcal{P}$ is also preserved by application of the unitary $U$, since the $Z$ terms only modify the coefficients in these linear combinations, and since the $XX$ phase flips both qubits at the same time. Moreover, applying any Pauli string on the pair of qubits also preserves property $\mathcal{P}$, since $Z$ terms only modifies the coefficients , and single $X$ terms applied on only one qubit of a pair map a linear combination of $|00\rangle$ and $|11\rangle$ into a linear combination of $|01\rangle$ and $|10\rangle$ and vice-versa. $Y$ terms can be seen as applying both $X$ and $Z$, up to an irrelevant global phase. Hence, property $\mathcal{P}$ holds true at all times. For each pair of qubits, which of the two types of linear combinations holds can be stored in memory using only one classical bit. The system can thus be effectively described by $L=20$ qubits and $L$ classical bits up to a global irrelevant phase. This technique allows thus for classical statevector simulation of these system sizes, whereas $N=40$ qubits with statevector simulation is usually completely out of reach with any reasonable resources. The precise update of the classical bits is performed as follows. Applying one $Z$ on any of the two qubits composing a pair is equivalent to applying a logical $Z$, up to an irrelevant global phase. Applying $X$ on the first qubit of a pair is equivalent to applying a logical $X$ and flipping the classical bit. Applying $X$ on the second qubit of a pair is equivalent to flipping the classical bit. Hence, when a depolarizing channel acts on two qubits that belong to different pairs, the Pauli operators can be taken to be logical operators, with each $X$ or $Y$ applied on one qubit automatically coming with a flipping of the classical bit of the pair to which the qubit belongs. As for depolarizing noise applied on the two qubits composing a pair, the effect is a bit less trivial. In Table \ref{tab:effective_noise}, we provide the probabilities of the effective noise channel applied on a pair of qubit (described by a qubit and a classical bit), with the Pauli operators acting on the (logical) qubit, and with $f$ the operator that flips the classical bit. 

\begin{table}[]
    \centering
    \begin{tabular}{|c|c|}
    \hline
    Error & probability\\
    \hline
       $I$  & $1-\frac{14\lambda}{15}$ \\
       $X$  & $2\lambda/15$\\
       $Y$  & $2\lambda/15$\\
       $Z$  & $2\lambda/15$\\
       $I f$  & $2\lambda/15$\\
       $X f$  & $2\lambda/15$\\
       $Y f$  & $2\lambda/15$\\
       $Z f$  & $2\lambda/15$\\
    \hline
    \end{tabular}
    \caption{Effective noise channel on a pair of qubit corresponding to depolarizing noise with amplitude $\lambda$.}
    \label{tab:effective_noise}
\end{table}

\subsection{Scaling of Pauli weight and Gate Count}
\label{sec_gate_count_and_weight}
\begin{figure*}[!ht]
	\centering
\includegraphics[width=1.0\textwidth]{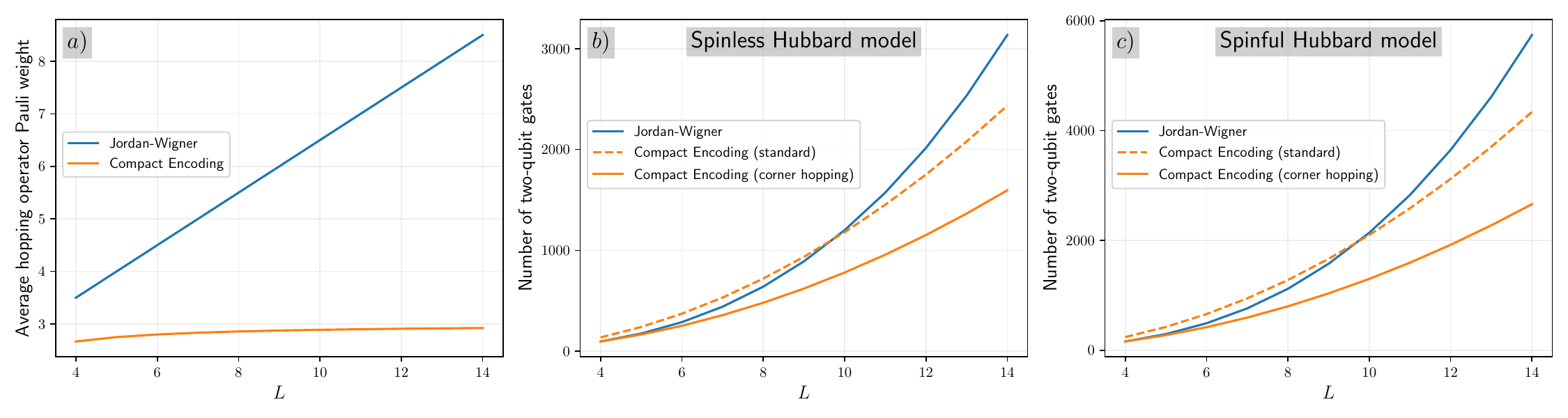}
\caption{\textbf{Average Pauli weight and two-qubit gate counts.}
a): Average weight of the hopping terms $c_i^\dagger c_j$ appearing in the Hamiltonian~\eqref{eq_hamiltonian}. b) Number of two-qubit gates $\exp(-i \theta ZZ)$ required for one Trotter step cf.~\eqref{eq_unitaries}. c) Same as b, but for the standard Hubbard model with both spin-up and spin-down fermions. All data for $L \times L$ models with open boundary conditions.
\label{fig_gatecounts}}
\end{figure*}
Our experimental data shows that the Compact Encodings compiled using Corner Hopping significantly outperforms the Jordan-Wigner encoding on a $6 \times 6$ lattice. How do we expect this advantage to scale to larger system sizes? And how important is the optimised compilation for this crossover to occur at $L<6$? We show in Fig.~\ref{fig_gatecounts} scalings of expected sensitivity (left and middle) and susceptibility to noise (right). We see that, for the parameters considered, the difference in average Pauli weight is small ($\sim 10\%$), which is compatible with the small, but significantly positive slope of the fit shown in Fig.~\ref{fig_weights}a. We expect that difference to increase on larger systems, for larger ratios of $t/U$, or for models with a larger fraction of hopping terms in the Hamiltonian (like e.g. a spinful Hubbard model with on-site interactions only). Regarding the susceptibility to noise (i.e. the number of gates), we see that the standard compilation not taking advantage of corner hopping would only show an advantage on much larger system sizes starting from $L \sim 10$. The exact gate count for the circuits run in this experiment is given in Table~\ref{tab_gate_count}.

\begin{table*}[h!]
    \centering
    \begin{tabular}{|c|c|c|c|c|c|}
     \hline
     \multicolumn{6}{|c|}{Two-Qubit Gate Count} \\
    \hline \hline
       &  State Preparation & 1 Trotter Step &  2 Trotter Steps &  3 Trotter Steps &  4 Trotter Steps \\
       \hline
    JW $4 \times 4$ & 0 & 54 & 108 & 162 & 216\\
    \hline
    CE $4 \times 4$ & 3 & 99 & 195 & 291 & 387 \\
    \hline
    JW $6 \times 6$ & 0 & 360 & 720 & 1080 & 1440\\
    \hline
    CE $6 \times 6$ & 12 & 264 & 516 & 768 & 1020 \\
    \hline
    \end{tabular}
    \caption{\textbf{Two-qubit gate count for all experiments}. All two-qubit gates are ZZPhase $(\exp(-i \theta/2 ZZ))$ which are native to the quantum computer that was used. JW stands for Jordan-Wigner and CE stands for Compact Encoding.}
    \label{tab_gate_count}
\end{table*}

\subsection{Effect of depolarizing noise on the fermionic encodings}
\begin{figure}
\centering
\includegraphics{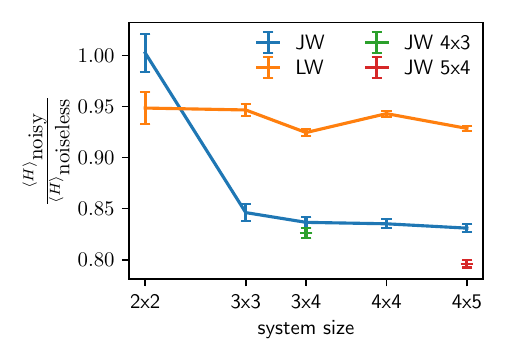}
\caption{Ratio of the noisy energy over the noiseless energy obtained by simulating the adiabatic algorithm circuit considered in the main text for a depolarizing noise with parameters $\lambda=0.003$ as function of system size for different geometries of the JW and LW encoding.}
\label{noisy_ratio}
\end{figure}

In this section, we study the effect of depolarizing noise on the energy measurement after an adiabatic time evolution as performed in the main text. We restrict ourselves to small system sizes such that the simulations can be performed exactly. We obtain the energy expectation value from the noiseless simulation of the circuit, and then simulate the effect of the depolarizing noise by simulating noisy trajectories. The results are shown on Fig. \ref{noisy_ratio}, where we plot the ratio $\langle H \rangle_\text{noisy}/\langle H \rangle_\text{noiseless}$.

The blue curve corresponds to the geometry yielding the shortest Pauli string for the JW Wigner encoding, while the green and red points (respectively 4x3 and 5x4) correspond to geometries with longer Pauli strings. The noisy over noiselss ratio is roughly independant of the system size for the LW encoding while it is strongly affected by the length of the Pauli strings for JW. The error bars represent the standard error of the mean associated with the average over the trajectories. Interestingly, for the 5x4 geometry, the maximum Pauli string length is 10, and the ratio is lower than in the 4x5 case, for which the maximum string length is 8, in accordance with the dilution of error picture~\cite{granet_dilution_2024}.

\subsection{Comparison with other compilation strategies}

\begin{figure*}[!ht]
	\centering
\includegraphics[width=1.0\textwidth]{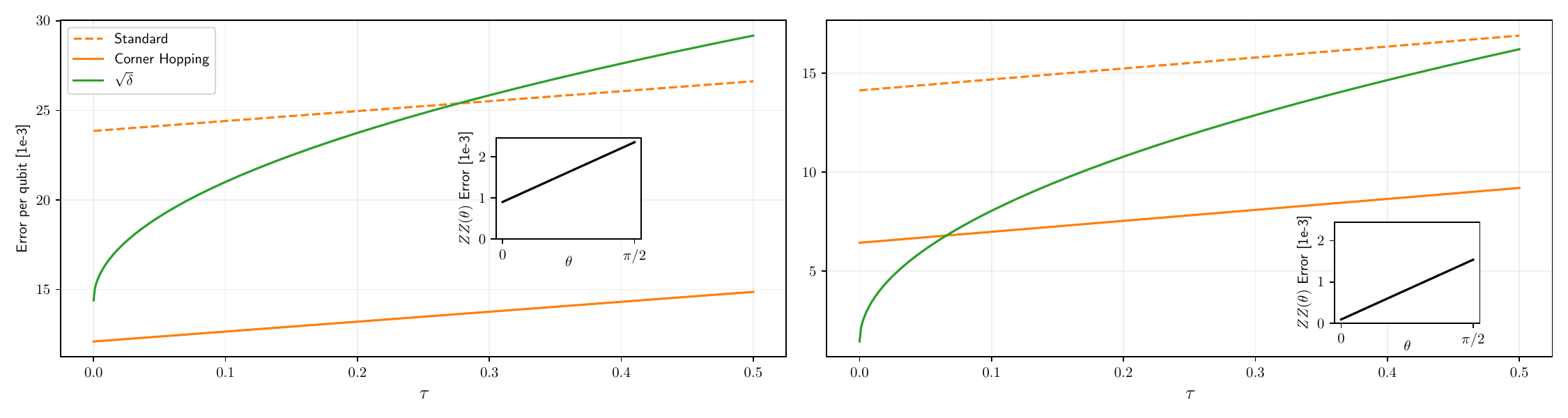}
\caption{\textbf{Gate error comparison for different compilation strategies.} Both panels show the average number of errors per qubit per Trotter step for nearest-neighbour fermionic hopping on the square lattice, as a function of the Trotters step size. Standard refers to Fig.~\ref{fig2}b before the optimisations, Corner Hopping refers to Fig.~\ref{fig2}b after the optimisations and $\sqrt{\delta}$ refers to the compilation strategy developed in~\cite{clinton_hamiltonian_2021}. We understand the latter decomposition to compile each 3-body operator into 4 gates of the form $\exp(-i\sqrt{\delta/2} P_i P_j P_k)$ with $P$ Pauli operators, i.e. using 16 partial entanglers per two bonds and no maximal entanglers.
a) Taking into account the current effective gate error profile for the native entangling gate $ZZ(\theta) = \exp(-i \theta/2 ZZ)$ (equation (1) in~\cite{moses_race-track_2023}). Even at zero Trotter step size, residual errors caused by wrapper pulses, laser phase noise and spontaneous emission remain (y-intercept in the inset). b) The $\sqrt{\delta}$ decomposition can be advantageous in a scenario where $dt$ is small and the constant offset in the gate error is reduced by a factor 10 (while the slope is kept constant). \label{fig_sqrt_delta}}
\end{figure*}

The corner hopping strategy developed in Fig.~\ref{fig2} showed good performance for the experiments reported in the main text. In Fig.~\ref{fig_sqrt_delta} we compare quantitatively the difference in gate error between three different compilation strategies, the standard strategy (i.e. Fig.~\ref{fig2} without using the circuit identities), the corner hopping, as well as the $\sqrt{\delta}$ strategy developed in \cite{clinton_hamiltonian_2021}. For the current status of the H2 quantum computer that was used in the present experiment, we see that the corner hopping strategy incurs the overall lowest error for all Trotter step sizes $\tau$, being roughly a factor 2 more efficient for the last Trotter step of the dynamics considered here ($\tau=0.2$). However, it is important to note that the $\sqrt{\delta}$-decomposition may be advantageous if the hardware noise in the quantum computer can be changed in such a way that noise on partial entanglers ($\tau \rightarrow 0$) is reduced. In that case, for the adiabatic path considered here, it would be advantageous to mix strategies, i.e., use the $\sqrt{\delta}$-decomposition in the beginning of the adiabtic evolution, where $\tau t(s)$ (the hopping parameter) is small and switch to the corner hopping strategy at a later point in the path.

\subsection{Effects of Trotter error}
\label{sec_trotter_error_encodings}

\begin{figure*}[!ht]
	\centering
\includegraphics[width=1.0\textwidth]{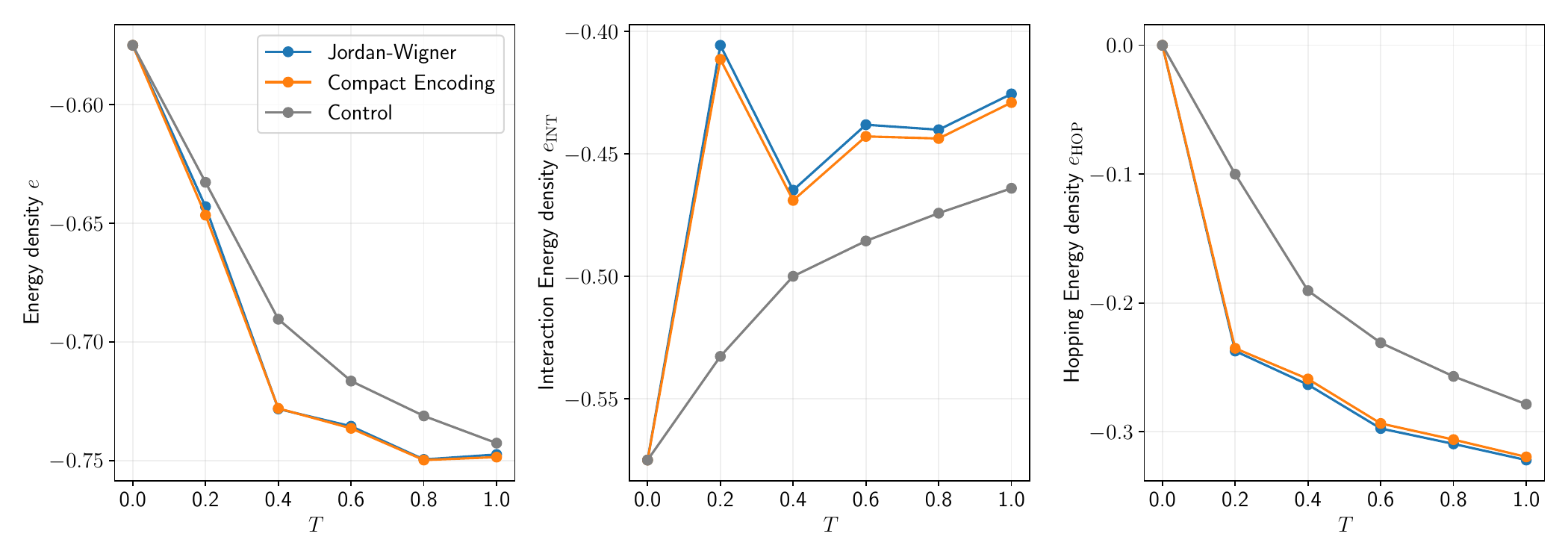}
\caption{\textbf{Effect of Trotter error in the different encodings.} Exact noiseless data for the adiabatic evolution~\eqref{eq_adiabatic_evolution} on the $4 \times 4$ system with $\tau = 0.2$, up to 5 Trotter steps. Control refers to a time series with much lower Trotter error, obtained using the Jordan-Wigner strategy with $\tau = 0.02$ and $T/\tau$ Trotter steps.
\label{fig_trotter_error}}
\end{figure*}

In the limit $\tau \rightarrow 0$, the compact encoding implemented via corner hopping and the Jordan-Wigner implemented via fermionic swap networks realise the same unitary evolution. In our scenario, where $\tau = 0.2$, the effect of Trotter errors is not a priori obvious---nor is it clear whether Trotter error affects both strategies in the same way. We provide noiseless data on the $4 \times 4$ system in Fig.~\ref{fig_trotter_error}. We make two observations: First, for the specific dynamics considered in this work, Trotter error is similar for both strategies and their difference is smaller than any of the error bars in the main text. Second, and maybe more surprisingly, the Trotterised adiabatic evolution with $\tau = 0.2$ reaches \textit{lower} energies at early times than a ``control" evolution implemented with a much smaller time step $\tau = 0.02$. To shed light on this phenomenon, we can expand the Floquet Hamiltonian $\exp(-i \tau H_F)$ implemented in one Trotter step
\begin{equation}
    H_F = H_0 + \tau t^2 \left(\sum_{\braket{\braket{ij}} \in \mathrm{corner}} (i c_i c_j + \mathrm{h.c.}) + \sum_{\braket{\braket{ij}} \notin \mathrm{corner}} (i c_i^\dagger c_j + \mathrm{h.c.})\right) + \tau \, U t \left(\sum_{\braket{ijk}} n_k(i c_i^\dagger c_j + \mathrm{h.c.})\right) + \mathcal{O}\left(\tau^2\right).
\end{equation}
Here $H_0$ is given in~\eqref{eq_hamiltonian}, next-nearest neighbour pairs $\braket{\braket{ij}}$ are distinguished by whether they are sitting at the corners of a triangle  implemented within the corner hopping scheme or not (and if there are two paths connecting $i$ and $j$ they need to be doubly counted) and $\braket{ijk}$ refers to triplets of neighbouring fermionic modes. The largest of these terms are the correlated hopping terms of the form $U t n_k(i c_i^\dagger c_j + \mathrm{h.c.})$. These lead the fermions to delocalise more quickly than in the continuous time evolution, at the cost of higher interaction energies. This picture is indeed born out in Fig.~\ref{fig_trotter_error}. The net effect of these Trotter errors is a ``Floquet cooling" of the system.

The fact that the complex amplitudes are purely imaginary can be physically interpreted as a magnetic field whose strength is of order $\tau$.

We also note that the same ideas can be used to implement next-nearest neighbour hopping. This $t'$-term has recently been shown to have a crucial influence on the superconductivity of the spinful Hubbard model~\cite{xu_coexistence_2024, ponsioen_period_2019}. In the compact encoding on the square lattice, it can be implemented using 3-body terms $(XY - YX)Z_\mathrm{ancilla}$ for hopping across an ancilla, whereas hopping across an empty plaquette would need to be carried out by a 5-body operator. Instead, real-valued next-nearest neighbour hopping can be implemented by exploiting Trotter error by conjugating the usual nearest-neighbour hopping $\braket{jk}$ with adjacent complex hopping terms on modes $\braket{kl}$:
\begin{equation}
\begin{aligned}
e^{i \tau_A (i c_k^\dagger c_l + \mathrm{h.c.})} \, e^{i \tau_B ( c_j^\dagger c_k + \mathrm{h.c.})} \, e^{-i \tau_A (i c_k^\dagger c_l + \mathrm{h.c.})} &= e^{i \tau_B (( c_j^\dagger c_k + \mathrm{h.c.}) + \tau_A ( c_l^\dagger c_j + \mathrm{h.c.})) + \mathcal{O}(\tau_A^2)} 
\end{aligned}
\end{equation}
which is a special case of the Baker-Campbell-Hausdorff formula at first order $\log \left( e^{i \, \tau_A A} e^{i \, \tau_B B} e^{-i \, \tau_A A}\right) /i\tau_B= B + i \, \tau_A [A,B] + \mathcal{O}(\tau_A^2)$.

\subsection{Parameters of the Adiabatic Evolution}
The adiabatic algorithm involves initializing the system in an easy to prepare ground state of Hamiltonian $H_0$ and then time evolving the state while slowly changing the Hamiltonian to the target Hamiltonian $H$. If the time evolution is sufficiently slow then the system remains in the ground state and its properties, such as energy, can be probed with measurements. 
For the spinless Hubbard model, the starting Hamiltonian contains only the repulsive interactions between the nearest-neighbours and the final Hamiltonian contains both hopping and repulsive interactions as described in Section II of the main text. 

There are many ways of implementing the ramp from $H_0$ to $H$ and it is important to design ramps that minimize the errors in the prepared state, given a finite amount of quantum computing resources. 
A digitized adiabatic algorithm has two sources of errors.
Firstly, there are non-adiabaticity errors due to the finite duration of the ramp, $T$. Secondly, there are Trotter errors due to the digitization of the adiabatic evolution into $n$ time intervals of duration $\tau=T/n$. 
In addition on a real quantum device, there are errors due to gate noise. The noise imposes a limit on the depth of the circuit that can be run on the device before all of the information is randomized, effectively fixing the maximum number of Trotter steps $n$. For a fixed $n$, there is a trade-off in the choice of $\tau$: a small $\tau$ results in small Trotter error but large non-adiabaticity error. 

\begin{figure}
    \centering
    \includegraphics[width=1.0\textwidth]{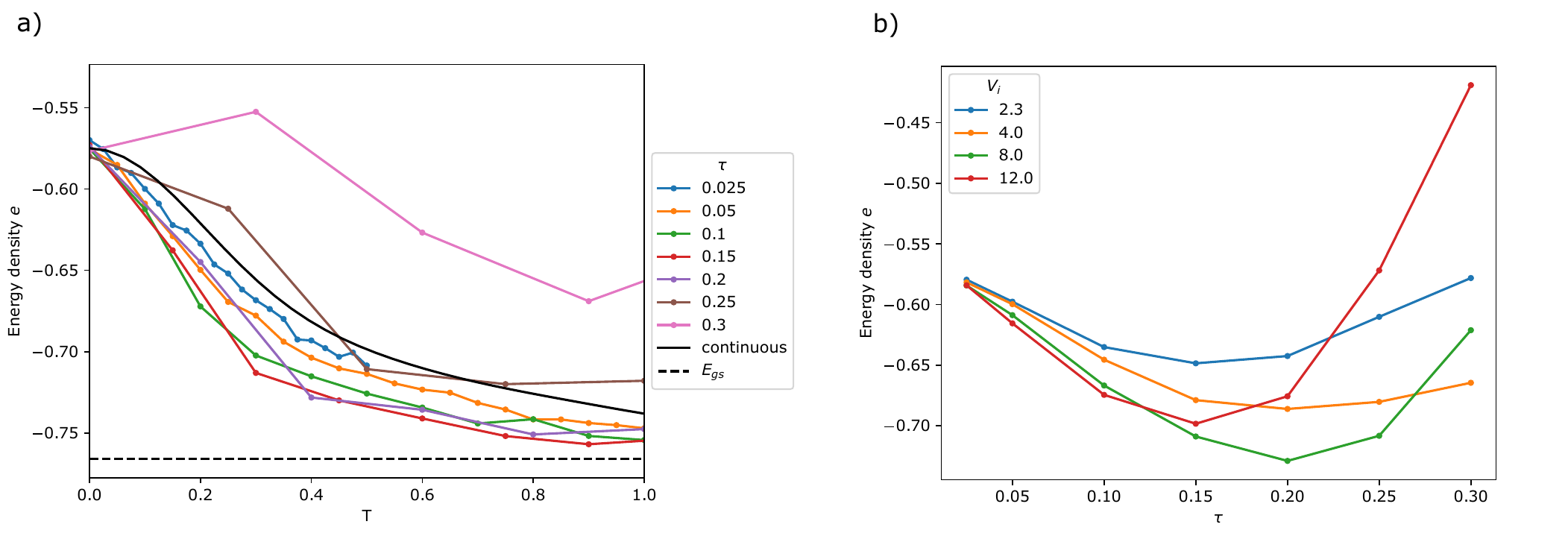}
    \caption{\textbf{Optimizing adiabatic path.} Noiseless Trotterized adiabatic evolution is simulated for 4x4 spinless Hubbard model with JW encoding. (a) Energy density in the end of the Trotterized adiabatic evolution for varying $\tau$ and total time $T$ with $V_i = 8.0$ and $V = 2.3$. The continuous black line corresponds to exact continuous evolution. (b) Energy density in the end of the ramp with 2 Trotter steps as a function of $\tau$ and $V_i$.}
    \label{fig:adiabatic_path_varying_dt_Ui}
\end{figure}

In order to select an optimal $\tau$ for the experiments reported in the main text, we have carried out a numerical simulations of the noiseless Trotterized circuits for 4x4 lattice. The results for various choices of $\tau$ and the total ramp time $T$ are shown in figure \ref{fig:adiabatic_path_varying_dt_Ui} a). An evenly discretized linear ramp in  $V$ and $t$ as shown in figure \ref{fig_main_result} a) was used. 
We have chosen to use $\tau=0.2$ in the experiments, as this results in a particularly rapid cooling, reducing the energy density to $-0.728$ in just two Trotter steps, where the system ground state energy is $-0.766$.

\begin{figure}
    \centering
    \includegraphics[width=1.0\textwidth]{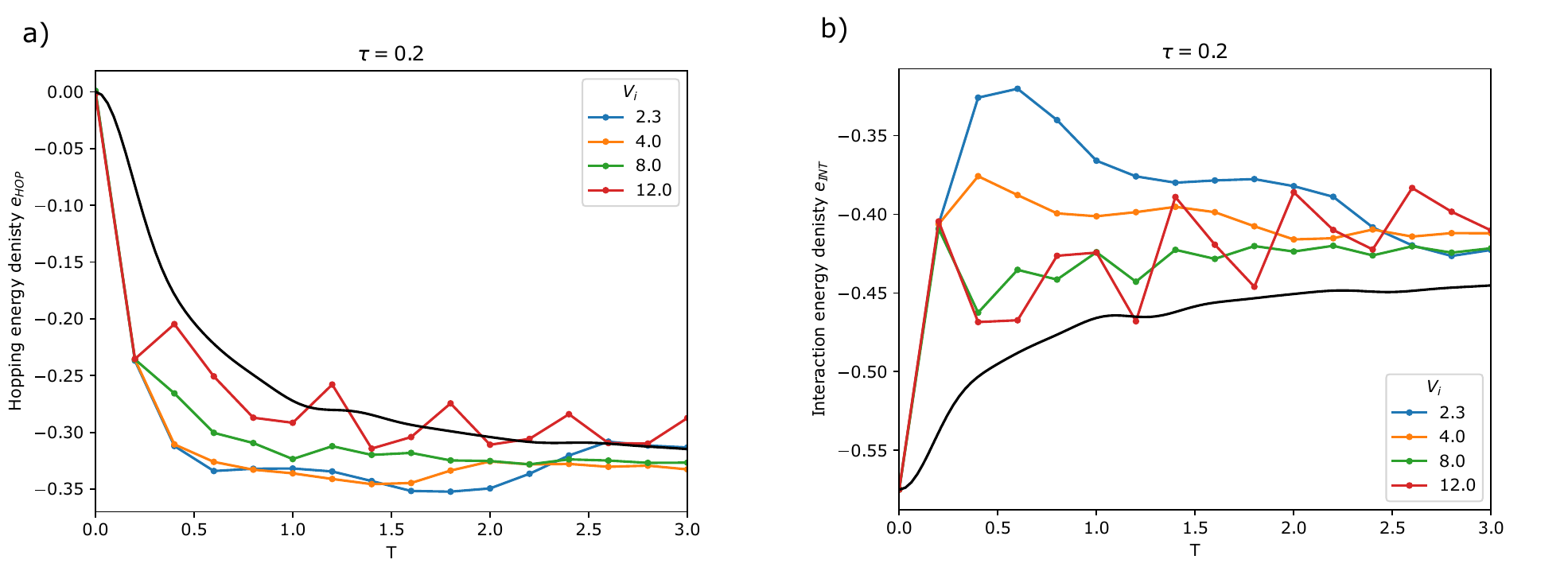}
    \caption{Hopping and interaction energy contributions of the state prepared with a Trotterized evolution with different $V_i$ and $T$. The time interval for each Trotter step is set to $\tau=0.2$. The continuous black line corresponds to exact continuous evolution.}
\label{fig:hopping_interaction_adiabatic_path}
\end{figure}

In addition to optimizing $\tau$, we have investigated varying the initial value of the interaction strength $V_i$. Figure \ref{fig:adiabatic_path_varying_dt_Ui} b) shows energy density after two Trotter steps as a function of $\tau$ and $V_i$. The lowest energy is obtained for $\tau \sim 0.2$ and $V_i \sim 8.0$, which are the parameters used in the H2 hardware experiments. We note that a small value of $V_i$ results in a faster convergence of the hopping energy, while a large value of $V_i$ results in a faster convergence of the interaction energy as can be seen from figure \ref{fig:hopping_interaction_adiabatic_path}. The choice, $V_i =8.0$, optimizes the sum of the two energy components.

\subsection{Non-scalability of global filtering}
\label{sec_non_scalability_particle_filtering}

We have shown that exponential shot overheads in system size coming from gate counts and sensitivity to error can be avoided by the compact encoding, as long as error rates remain constant with increasing qubit number. However, as long as noise is not negligible, a final step is usually carried out in which error mitigation is applied to the data. One has to be careful to avoid introducing exponential dependencies back into the process in this final step.

\begin{figure*}[!ht]
	\centering
\includegraphics[width=1.0\textwidth]{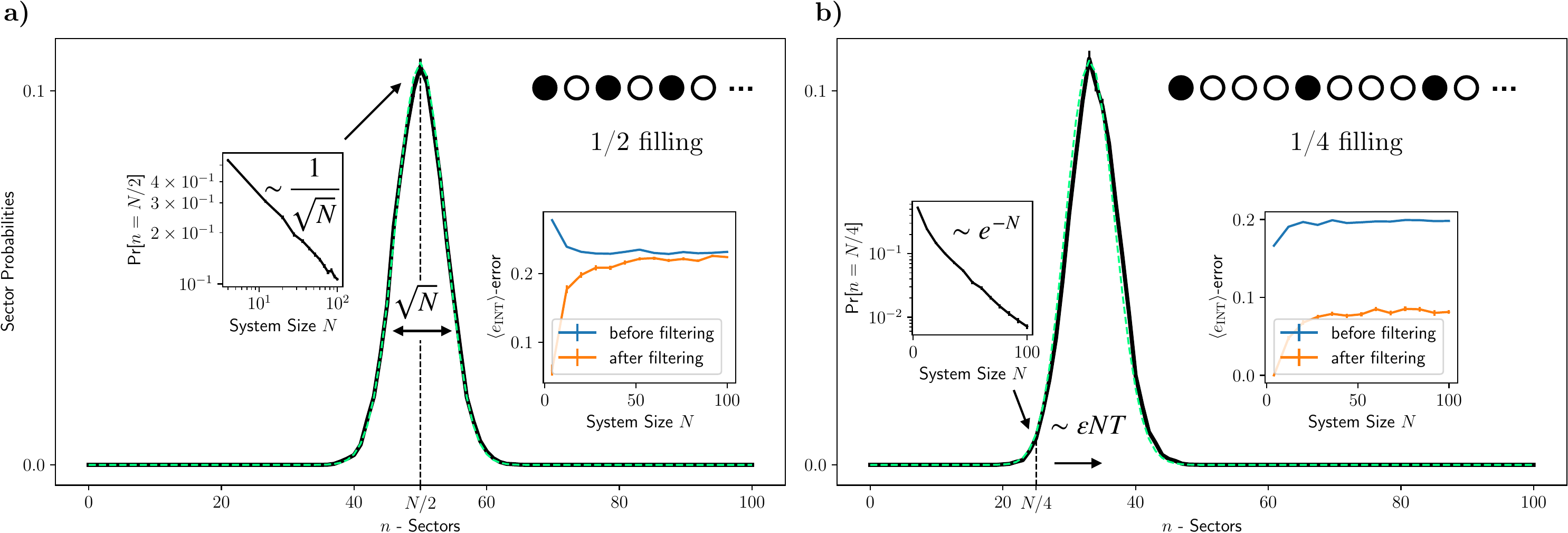}
\caption{\textbf{Issues with scaling particle number filtering to larger system sizes.} An $N$-site Hubbard chain is evolved for three noisy Trotter steps, starting from two different initial states: a) For the half-filled state, noise causes the particle number $n$ to carry out a random walk, which diminishes the fraction of retained shots to $1/\sqrt{N}$ (black inset). The error on the interaction energy density $e_\mathrm{INT} = V \sum_i^{N-1} n_i n_{i+1}/(N-1)$ of the retained shots approaches that of the raw data in the thermodynamic limit (orange/blue inset). b) For the quarter-filled state, $n$ walks towards $N/2$, covering a distance of $N (e^{-\varepsilon T }-1)/2 = \mathcal{O}( \varepsilon N T)$. The retained shots contain more useful information than in the half-filled case (orange/blue inset), but all but an exponentially small fraction of shots have to be discarded. Error bars are standard errors on the mean originating from shot noise. Dashed green shows predictions from the random walk model~\eqref{eq_random_walk}.
\label{fig_scalability1}}
\end{figure*}

One commonly used technique for Hubbard model and general quantum chemistry simulations is to filter shots by symmetry: if one initialises the system in a state in a known particle sector $n \ket{\psi} := \sum_i n_i \ket{\psi} = n_0 \ket{\psi}$, and the particle number commutes with the desired evolution $U$ as well as with the observable of interest $M$, i.e., $[n,U]=[n,M]=0$, then one may measure simultaneously $M$ and $n$ and throw away all shots in which $\Delta n = |n-n_0| > 0$. The shots discarded in this way are guaranteed to include errors, either from the noise in the quantum computer or from symmetry-breaking terms in the Trotterisation. While symmetry filtering works well for the system sizes we have considered in the experiment, here we show that post-selecting on global constraints does come with exponential overheads in system size and is thus a band-aid solution for the current moment at best. To see this, we classically simulate the Hubbard model~\eqref{eq_hamiltonian}, again with parameters $t=1$ and $V=2.3$, albeit on a one-dimensional chain geometry with $N$ sites. We keep $V$ and $t$ fixed and execute three Trotter steps with step size $\tau=0.1$ on top of an initial product state at half-(quarter-)filling shown in Fig.~\ref{fig_scalability1}a (b), depolarising each qubit with probability $\varepsilon = 10\%$ after each Trotter step. We then evaluate the probability to find the quantum state in a given number sector.

In both scenarios, the effect of errors on the particle number can be described by a biased random walk on the line $0, \dots N$ with transition probabilities
\begin{equation}
\label{eq_random_walk}
\begin{aligned}
    \mathrm{Prob}[n \rightarrow n-1] &= \varepsilon' \frac{n}{N} \\
    \mathrm{Prob}[n \rightarrow n] &= 1-\varepsilon' \\
    \mathrm{Prob}[n \rightarrow n+1] &= \varepsilon' \frac{N-n}{N},
\end{aligned}
\end{equation}
where $\varepsilon' = 2/3 \varepsilon$, since only the $X$- and $Y$-components of the depolarising error change the particle number but $Z$ does not. In the half-filled scenario, the initial state is in the infinite temperature sector of the observable $n$ and thus, the mean stays at $n=N/2$. The width (height) of the distribution increases (decreases) to a value proportional to $\sqrt{N}$ ($1/\sqrt{N}$) but independent of $\varepsilon$, to match that of the steady state distribution $\mathrm{Pr}[n=k] \propto \binom{N}{k}$, cf. the loglog inset in Fig.\ref{fig_scalability1}a. Since the height is exactly the probability for a given shot to be measured in the desired particle sector, one will retain $1/\sqrt{N}$ shots after particle number filtering. However, the random walk picture also predicts that the utility of the filter will decrease with system size: Almost all events in which the walker finds itself at $n=N/2$ are not due to the absence of errors but are rather due to a coincidental return to the initial position. To make this notion concrete, we compute the error on the expectation value of the interaction energy density $e_\mathrm{INT} = V\sum_i n_i n_{i+1}/(N-1)$, before (blue) and after filtering (orange). The inset in Fig.~\ref{fig_scalability1}a shows that the error before and after filtering converges with system size to the same value. We conclude that, at half-filling, particle number filtering will retain a polynomial fraction of the noisy data, but has vanishing effect at large $N$.

Filtering at quarter-filling $n_0 = N/4$ also fails for large $N$, but for a different reason: Unlike in the half-filled case, the bias causes a departure from $n_0$ towards the infinite temperature value $n_\infty = N/2$, in particular the mean $n_T$ moves in time $T$ according to
\begin{equation}
    \frac{n_T-n_\infty}{n_0-n_\infty} \sim e^{-\varepsilon T}
\end{equation}
Since the distance between the mean and $n_0$ is extensive in $N$, but the distribution only has width $\sqrt{N}$, the probability of finding a shot in the initial sector decreases as $e^{-N}$ in system size, as confirmed in the semilogarithmic inset in Fig.~\ref{fig_scalability1}b. The net effect of the filtering does not vanish in the thermodynamic limit (cf. the inset in the same figure), but in order to carry it out, all but an exponentially small fraction of the shots must be discarded. One idea to circumvent this problem is to use a ``soft" filter i.e instead of projecting the raw density matrix using $\rho \rightarrow P \rho P$ with $P = \delta_{n,n_0}$, for example, for $n_0 < n_\infty$ one may choose 
\begin{equation}
    P \propto \begin{cases}
			1, & \text{if $n<n_T$}\\
                -an + b, & \text{if $n_T < n < n_T+\sqrt{N}$}\\
                0, & \text{if $n > n_T+\sqrt{N}$},
		 \end{cases}
\end{equation}
where $2/a = \sqrt{N}$ is the expected width of the distribution and $b = -a(n_T + \sqrt{N}/2)$.

\subsection{Random walk model for errors in non-trivial quantum dynamics}
\begin{figure*}[!ht]
	\centering
\includegraphics[width=1.0\textwidth]{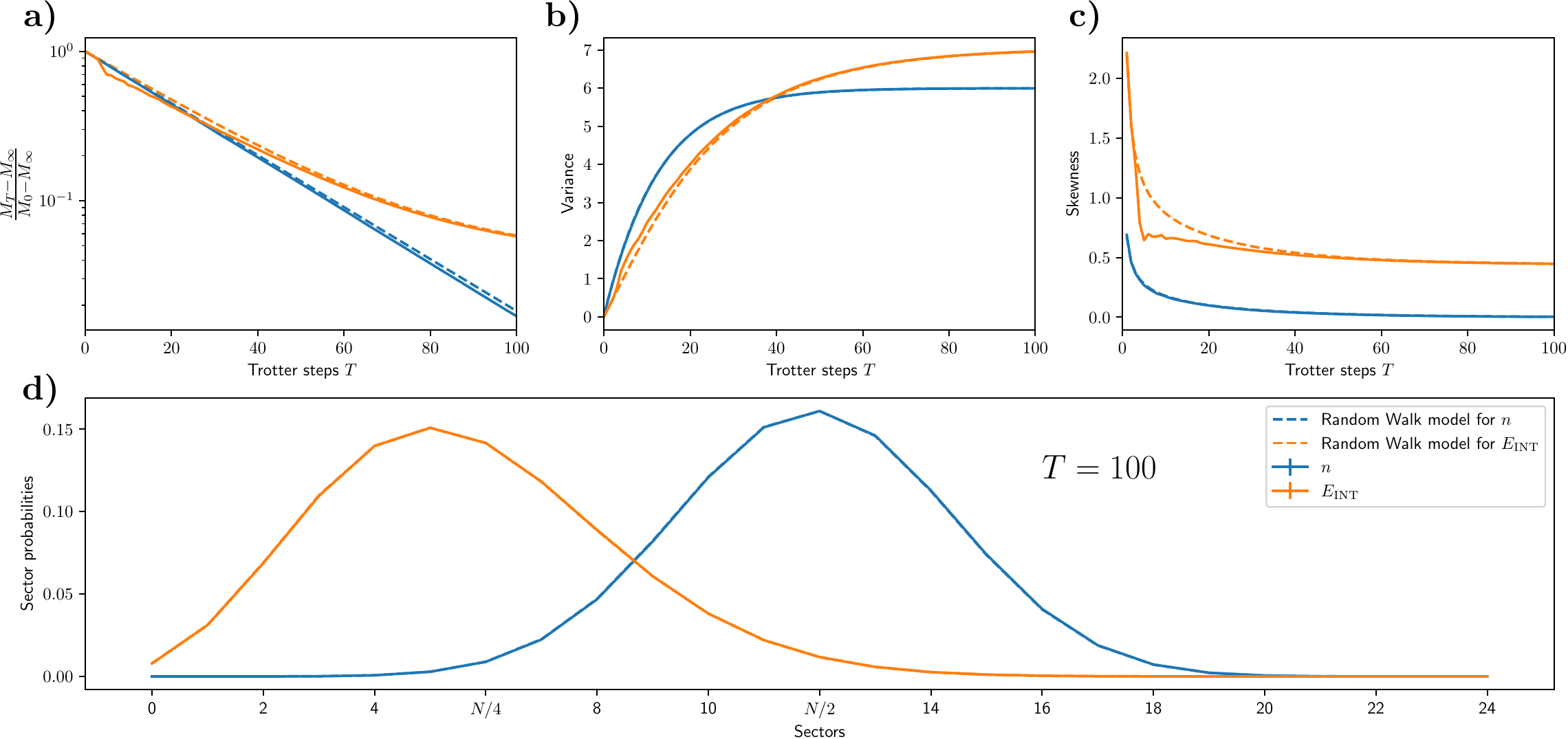}
\caption{\textbf{Random walk model for trivial and non-trivial noisy dynamics.} A 24-site Hubbard chain is evolved for up to 100 noisy $\tau=0.1$ Trotter steps, starting from the quarter-filled initial state shown in Fig.~\ref{fig_scalability1}b and measuring the distributions of both the (conserved) particle number $n=\sum_{i=1}^{N} n_i$ as well as the (non-conserved) interaction energy $E_\mathrm{INT} = \sum_{i=1}^{N-1} n_i n_{i+1}$. The time evolution of the mean, variance and skewness of these distributions is shown in a, b and c and a snapshot after $T=100$ Trotter steps is presented in d. The random walk model~\eqref{eq_random_walk} captures almost perfectly the distribution of the particle number $n$. Much more surprising is that the distribution of $E_\mathrm{INT}$ is captured by a similar random walk model that runs in polynomial time on a classical computer, up to a transient regime up to roughly 30 Trotter steps where quantum dynamics is relevant. Data is superimposed when only solid lines are visible. In a, $n_\infty = N/2$, ${E_\mathrm{INT}}_\infty = N/4$, $n_0 = N/4$, ${E_\mathrm{INT}}_0 = 0$.
\label{fig_random_walk_vs_hubbard}}
\end{figure*}
Motivated by the success of a simple random walk model for the effect of errors on the particle number in section~\ref{sec_non_scalability_particle_filtering}, we now would like to understand whether (i) this model can also capture the long-time dynamics and (ii) check its predictions on an observable that undergoes non-trivial quantum dynamics and is hard to compute classically from first principles.

To assess the long-time effects of errors, we initialise a 24-site Hubbard chain~\eqref{eq_hamiltonian} with $t=1$ and $V=2.3$ in the quarter-filled product state shown in Fig.~\ref{fig_scalability1}b. Then we evolve the system for $T=1,2, \dots 100$ first-order Trotter steps with $\tau = 0.1$, applying a single-qubit depolarising channel with $\varepsilon=3\%$ to each qubit after each Trotter step. After each step, we measure the distribution of the particle number $n=\sum_{i=1}^{N} n_i$ as well as the interaction energy $E_\mathrm{INT} = \sum_{i=1}^{N-1} n_i n_{i+1}$. Since the particle number is conserved by the quantum dynamics, we expect the random walk model~\eqref{eq_random_walk} to accurately capture its distribution for all times and this is indeed confirmed in Fig.~\ref{fig_random_walk_vs_hubbard}.

Much more surprising is the effect of errors on the interaction energy $E_\mathrm{INT}$. Unlike the particle number, this observable is not conserved and undergoes non-trivial quantum dynamics, a situation which generally is notoriously difficult to capture with classical methods. We nevertheless attempt to capture at least the effect of errors on $E_\mathrm{INT}$ with a classical random process: We initialise a bitstring $b_0 = [1,0,0,0,1,0,0,0,\dots]$ and, in each step, flip each bit with probability $2\%$. This model neglects the quantum dynamics entirely and might be considered appropriate if the quantum circuit contains only error processes but no gates.
The results of this process are shown in the dashed orange lines of Fig.~\ref{fig_random_walk_vs_hubbard}. Two regimes can be identified. There is an initial, transient regime up to roughly 30 Trotter steps, in which the classical model fails to capture the quantum dynamics. From then on, however, the distribution of the classical model captures not only the mean but becomes effectively indistinguishable from that of the quantum dynamics, i.e. the dynamics do not alter the distribution at late times, which is almost perfectly captured by a classical model.

\end{document}